\documentclass[aps,prl,reprint,longbibliography, superscriptaddress]{revtex4-1}

 \usepackage[utf8]{inputenc}
 \usepackage[T1]{fontenc}
 \usepackage{amsmath,bm}
 \usepackage{mathrsfs}
 \usepackage{amsfonts}
 \usepackage{graphicx}
 \usepackage{setspace} 
 \usepackage{graphicx}
 \usepackage{epstopdf}
 \usepackage{dcolumn}
 \usepackage{amsmath}
 \usepackage{epsfig}
 \usepackage{indentfirst}
 \usepackage{psfrag}
 \usepackage{subfigure}
 \usepackage{amssymb}
 \usepackage{color}
 \usepackage{units} 
 \usepackage{gensymb}
 \usepackage{xr}
\makeatletter
\newcommand*{\addFileDependency}[1]{
  \typeout{(#1)}
  \@addtofilelist{#1}
  \IfFileExists{#1}{}{\typeout{No file #1.}}
}
\makeatother

\newcommand*{\myexternaldocument}[1]{%
    \externaldocument{Appendix}%
    \addFileDependency{Appendix.tex}%
    \addFileDependency{Appendix.aux}%
}
\myexternaldocument{Appendix}

 \usepackage{graphicx}
 \usepackage{dcolumn}
 \usepackage{bm}
 \usepackage{natbib}
\usepackage{multirow}

\usepackage{physics}
\usepackage{dcolumn}
\usepackage{bm}
\usepackage{natbib}
\usepackage[backref=none,bookmarksnumbered=true,bookmarks=true,bookmarksopen=true,colorlinks=true,
citecolor=blue,linkcolor=blue,anchorcolor=green,urlcolor=blue,unicode=false]{hyperref}
\usepackage{orcidlink}

\usepackage{ulem}[normalem]

\makeatletter
\newcommand\colorsout[1]{\bgroup \markoverwith{\textcolor{#1}{\rule[0.5ex]{2pt}{0.4pt}}}\ULon}

\makeatother

\newcommand{\beginsupplement}{%
        \setcounter{table}{0}
        \renewcommand{\thetable}{S\arabic{table}}%
        \setcounter{figure}{0}
        \renewcommand{\thefigure}{S\arabic{figure}}%
     }

\begin{document}
\title{Epitaxial monolayers of magnetic 2D semiconductor FeBr\textsubscript{2} grown on Au(111)}

\author{S. E. Hadjadj \footnote{Corresponding first author} \orcidlink{0000-0002-6045-574X}}
  \thanks{These two authors contributed equally}
  \affiliation{Freie Universität Berlin, Institut für Experimentalphysik, Arnimallee 14, 14195 Berlin, Germany}

\author{C. Gonz\'alez-Orellana\footnote{Corresponding first author} \orcidlink{0000-0003-4033-5932}}
  \thanks{These two authors contributed equally}
  \affiliation{Centro de F\'isica de Materiales (CSIC/UPV-EHU), 20018 Donostia-San Sebasti\'an, Spain}

\author{J. Lawrence \orcidlink{}}
      \affiliation{Donostia International Physics Center, 20018 Donostia-San Sebasti\'an, Spain}

\author{D. Bikaljevi\'c \orcidlink{0000-0003-3409-7042}}
\affiliation{CIC nanoGUNE-BRTA, 20018 Donostia-San Sebasti\'an, Spain}
\affiliation{Institute of Physical Chemistry, University of Innsbruck, Innrain 52c, A-6020 Innsbruck, Austria}

\author{M. Peña-D\'iaz \orcidlink{}}
 \affiliation{Centro de F\'isica de Materiales (CSIC/UPV-EHU), 20018 Donostia-San Sebasti\'an, Spain}
 
\author{P. Gargiani \orcidlink{0000-0002-6649-0538}}
    \affiliation{ALBA Synchrotron Light Source, 08290 Barcelona, Spain} 
 
\author{L. Aballe \orcidlink{0000-0003-1810-8768}}
        \affiliation{ALBA Synchrotron Light Source, 08290 Barcelona, Spain} 
        
\author{J. Naumann \orcidlink{0000-0003-4513-5139}}
  \affiliation{Freie Universität Berlin, Dahlem Center for Complex Quantum Systems, Arnimallee 14, 14195 Berlin, Germany}    
  
\author{M. \'A. Niño \orcidlink{0000-0003-3692-147X}}
        \affiliation{ALBA Synchrotron Light Source, 08290 Barcelona, Spain} 
        
\author{M. Foerster \orcidlink{0000-0002-4147-6668}}
    \affiliation{ALBA Synchrotron Light Source, 08290 Barcelona, Spain}
    
\author{S. Ruiz-G\'omez \orcidlink{}}
    \affiliation{Max Planck Institute for Chemical Physics of Solids, 01180, Dresden, Germany}
 
\author{S. Thakur \orcidlink{0000-0003-4879-5650}}
  \affiliation{Freie Universität Berlin, Institut für Experimentalphysik, Arnimallee 14, 14195 Berlin, Germany}
  
\author{I. Kumberg \orcidlink{0000-0002-3914-0604}}
  \affiliation{Freie Universität Berlin, Institut für Experimentalphysik, Arnimallee 14, 14195 Berlin, Germany}
    
\author{J. Taylor \orcidlink{0000-0001-5274-8545}}
    \affiliation{Helmholtz-Zentrum  Berlin, Albert-Einstein-Str. 15, 12489 Berlin, Germany}
    \affiliation{Fakultät für Physik, Technische Universität München, James-Franck-Straße 1, 85748 Garching bei München, Germany}

\author{J. Hayes \orcidlink{0000-0002-2740-1213}}
    \affiliation{Freie Universität Berlin, Institut für Experimentalphysik, Arnimallee 14, 14195 Berlin, Germany}

\author{J. Torres \orcidlink{0009-0005-6767-3757}}
    \affiliation{Freie Universität Berlin, Institut für Experimentalphysik, Arnimallee 14, 14195 Berlin, Germany}

\author{C. Luo \orcidlink{0000-0001-6476-9116}}
    \affiliation{Helmholtz-Zentrum  Berlin, Albert-Einstein-Str. 15, 12489 Berlin, Germany}
    \affiliation{Fakultät für Physik, Technische Universität München, James-Franck-Straße 1, 85748 Garching bei München, Germany}

\author{F. Radu \orcidlink{0000-0003-0284-7937}}
    \affiliation{Helmholtz-Zentrum  Berlin, Albert-Einstein-Str. 15, 12489 Berlin, Germany}

\author{D. G. de Oteyza \orcidlink{0000-0001-8060-6819}}
    \affiliation{Donostia International Physics Center, 20018 Donostia-San Sebasti\'an, Spain}
    \affiliation{Nanomaterials and Nanotechnology Research Center (CINN), CSIC-UNIOVI-PA, Oviedo, Spain}

\author{W. Kuch \orcidlink{0000-0002-5764-4574}}
  \affiliation{Freie Universität Berlin, Institut für Experimentalphysik, Arnimallee 14, 14195 Berlin, Germany}

\author{J. I. Pascual \orcidlink{0000-0002-7152-4747}}
\affiliation{CIC nanoGUNE-BRTA, 20018 Donostia-San Sebasti\'an, Spain}
\affiliation{Ikerbasque, Basque Foundation for Science, 48013 Bilbao, Spain}
 
\author{C. Rogero \orcidlink{0000-0002-2812-8853}}
 \affiliation{Centro de F\'isica de Materiales (CSIC/UPV-EHU), 20018 Donostia-San Sebasti\'an, Spain}
    \affiliation{Donostia International Physics Center, 20018 Donostia-San Sebasti\'an, Spain}
 
 \author{M. Ilyn\footnote{Corresponding first author} \orcidlink{0000-0002-4052-7275}}
 \thanks{Corresponding author}
  \affiliation{Centro de F\'isica de Materiales (CSIC/UPV-EHU), 20018 Donostia-San Sebasti\'an, Spain}

\begin{abstract}

\noindent Magnetic two-dimensional (2D) semiconductors have attracted a lot of attention because modern preparation techniques are capable of providing single crystal films of these materials with precise control of thickness down to the single-layer limit. It opens up a way to study rich variety of electronic and magnetic phenomena with promising routes towards potential applications. We have investigated the initial stages of epitaxial growth of the magnetic van der Waals semiconductor FeBr\textsubscript{2} on a single-crystal Au(111) substrate by means of low-temperature scanning tunneling microscopy, low-energy electron diffraction, x-ray photoemission spectroscopy, low-energy electron emission microscopy and x-ray photoemission electron microscopy. Magnetic properties of the one- and two-layer thick films were measured via x-ray absorption spectroscopy/x-ray magnetic circular dichroism. Our findings show a striking difference in the magnetic behaviour of the single layer of FeBr\textsubscript{2} and its bulk counterpart, which can be attributed to the modifications in the crystal structure due to the interaction with the substrate.
\end{abstract}

\date{\today}

\maketitle

\section{Introduction}
\noindent Integration of two-dimensional (2D) materials in technologically relevant applications requires atomic-scale control of the growth of single crystalline, monolayer-thick films. Meanwhile many semiconducting 2D materials like graphene, h-BN or MoS\textsubscript{2} are routinely grown on the wafer scale~\cite{zhang_strategies_2021}, preparation of magnetic 2D materials is still limited in most cases to micromechanical exfoliation ~\cite{huang_layer-dependent_2017, sethulakshmi_magnetism_2019, ashton_two-dimensional_2017, li_intrinsic_2019, gibertini_magnetic_2019, cortie_twodimensional_2020, li_2d_2021, zhao_structure_2021,huang_twodimensional_2021}. Prominent exceptions of this trend are magnetic transition-metal tri- and dihalides, for which single-layer growth was demonstrated recently via molecular beam epitaxy~\cite{kezilebieke_topological_2020, bikaljevic_noncollinear_2021}. In contrast to well-studied trihalides, particularly CrI\textsubscript{3} and CrBr\textsubscript{3} \cite{chen_direct_2019,huang_layer-dependent_2017,huang_electrical_2018,bonilla_strong_2018}, experimental investigation of the 2D dihalides is less advanced, although their bulk magnetic properties were thoroughly studied~\cite{mcguire_crystal_2017}.     

\noindent Bulk FeBr\textsubscript{2} is a layered crystal that consists of covalently bonded layers stacked via van der Waals (vdW) interactions in the CdI\textsubscript{2}-type structure (P3m1 space group). The layers consist of triangular lattices of cations in edge-sharing octahedral coordination 1T (or D\textsubscript{3d})- MX\textsubscript{2} structure, forming one transition metal layer sandwiched between two halide layers~\cite{mcguire_crystal_2017, chhowalla_chemistry_2013}. The lateral lattice constant was found to be 3.776 \text{ \AA} \cite{haberecht_refinement_2001,youn_large_2002}. Indirect Fe-Fe exchange interaction gives rise to the collinear intralayer ferromagnetic order below $T_{N}=14.2$~K with out of plane (OOP) anisotropy, meanwhile the interlayer exchange is antiferromagnetic. Application of an external magnetic field of 3.15~T triggers a metamagnetic phase transition~\cite{mcguire_crystal_2017,yang_van_2021}. The six 3d electrons of the Fe\textsuperscript{2+} ions are distributed between two groups of orbitals, t$_{2g}$ (d$_{xy}$, d$_{xz}$ and d$_{yz}$) and e$_{g}$ (d$_{x^{2} - y^{2}}$, and d$_{z^{2}}$), giving rise to a magnetic moment of 4.4 $\pm$ 0.7 $\mu_{B}/$Fe atom \cite{wilkinson_neutron_1959}, which exceeds the value of 4.0 $\mu_{B} /$ Fe atom predicted by Hund's rule~\cite{ropka_electronic_2001,pelloth_local_1995}. Various attempts of DFT calculations yield comparable values of the magnetic moments and provide useful insights on the details of the band structure~\cite{kulish_single-layer_2017, botana_electronic_2019, sargolzaei_spin_2008}.

\noindent In this work we use sublimation of the stoichiometric powder to grow epitaxial films of magnetic semiconductor FeBr\textsubscript{2}, which belongs to the family of transition metal dihalides (TMDH)~\cite{mcguire_crystal_2017}, on the single crystal Au(111). Feasibility of growth of TMDH films via Chemical Vapor Deposition (CVD) has been demonstrated recently~\cite{Jiang_2023}. In contrast to CVD, Molecular Beam Epitaxy (MBE) does not require heating of the substrate above room temperature which makes it compatible with resist-based nanofabrication and opens up a way to the integration of the TMDH thin films in the scalable manufacturing processes. We focus our investigation on the properties of the one- and two-slab thick films employing spectroscopic and microscopic characterisation, including synchrotron-based techniques. In particular, we demonstrate the modification of the magnetic properties of the stoichiometric FeBr\textsubscript{2} due to a reconstruction in the first slab.  

\section{Experiment and methods} 

\noindent FeBr\textsubscript{2} layers with variable thicknesses, ranging from sub-monolayer (sub-ML) to more than one monolayer, were grown on Au(111) using FeBr\textsubscript{2} powder from Sigma Aldrich with a purity of 98\% and a Knudsen cell evaporator. The sublimation temperature for FeBr\textsubscript{2} was around 400 $^{\circ}$C in ultra high vacuum (UHV) (with an evaporation pressure of 10\textsuperscript{-8} mbar to 10\textsuperscript{-9} mbar). The substrate was kept at room temperature during sublimation. A quartz microbalance was used to measure the nominal amount of the deposited material, meanwhile the calibration of the absolute thickness was done via cross-correlation of scanning tunneling microscopy (STM) images with low energy electron diffraction (LEED) data. This calibration was translated to the integral of the non-polarized soft X-ray absorption at the Fe L\textsubscript{3,2} edges for comparison to samples prepared in different synchrotron radiation sources. The thickness calibration procedure is shown in the supplementary information in Fig. \ref{fig:MagicAngle}. The Au(111) substrate was cleaned by standard Ar\textsuperscript{+} sputtering and annealing cycles.
Low-temperature STM (LT-STM) experiments were performed at 4.3~K (for a sub-ML sample) and at 77~K for the thicker samples at Centro de Física de Materiales and BOREAS beamline, respectively.

\noindent X-ray photoelectron spectroscopy (XPS) measurements were carried out with a Phoibos 100 photoelectron spectrometer, using a non-monochromatic Al-K$\alpha$ X-ray source. The analyser energy resolution is 0.1~eV. UHV was preserved during all the sample transfers (base pressure during experiment was 10\textsuperscript{-10}~mbar). 

\noindent X-ray magnetic circular dichroism (XMCD) measurements were performed at both the VEKMAG station (dipole-beamline) of BESSY~II in Berlin \cite{noll_mechanics_2017} and the BOREAS beamline (undulator-beamline) at ALBA Synchrotron Light Facility \cite{barla_design_2016}. The measurements at VEKMAG were performed by keeping the beam polarization constant and changing the field. At BOREAS we kept the field constant and changed the polarization.
Absorption spectra at the Fe L\textsubscript{3,2}-edges were acquired at normal incidence (NI/$0^\circ$, out of plane) and grazing incidence (GI/$70^\circ$, in plane), applying a variable magnetic field up to $\pm$6~T. The temperature during the measurements at the VEKMAG beamline was set to 10~K, which is around $12.6$~K and at BOREAS $2\pm 0.5$~K. One 0.6-ML sample of FeBr\textsubscript{2} was brought by a Ferrovac suitcase to BOREAS beamline to cross-correlate the coverage of the samples measured in the home laboratory and in the synchrotron beamlines. A control sample with similar coverage was grown in-situ and characterized at the BOREAS beamline. XAS/XMCD measurements did not show substantial differences between the samples.

\noindent The LEED images were acquired to observe the growth and the thickness-dependent change in the structure. 
Imaging at the mesoscopic scale was done by low-energy electron microscopy (LEEM) and x-ray photoemission electron microscopy (XPEEM) at the CIRCE beamline (ALBA Synchrotron Light Facility) \cite{aballe_alba_2015}.

\section{Results and discussion}
\subsection{Epitaxial Growth of FeBr\textsubscript{2}}

\begin{figure*}[ht!]
\includegraphics[width=0.9\textwidth]{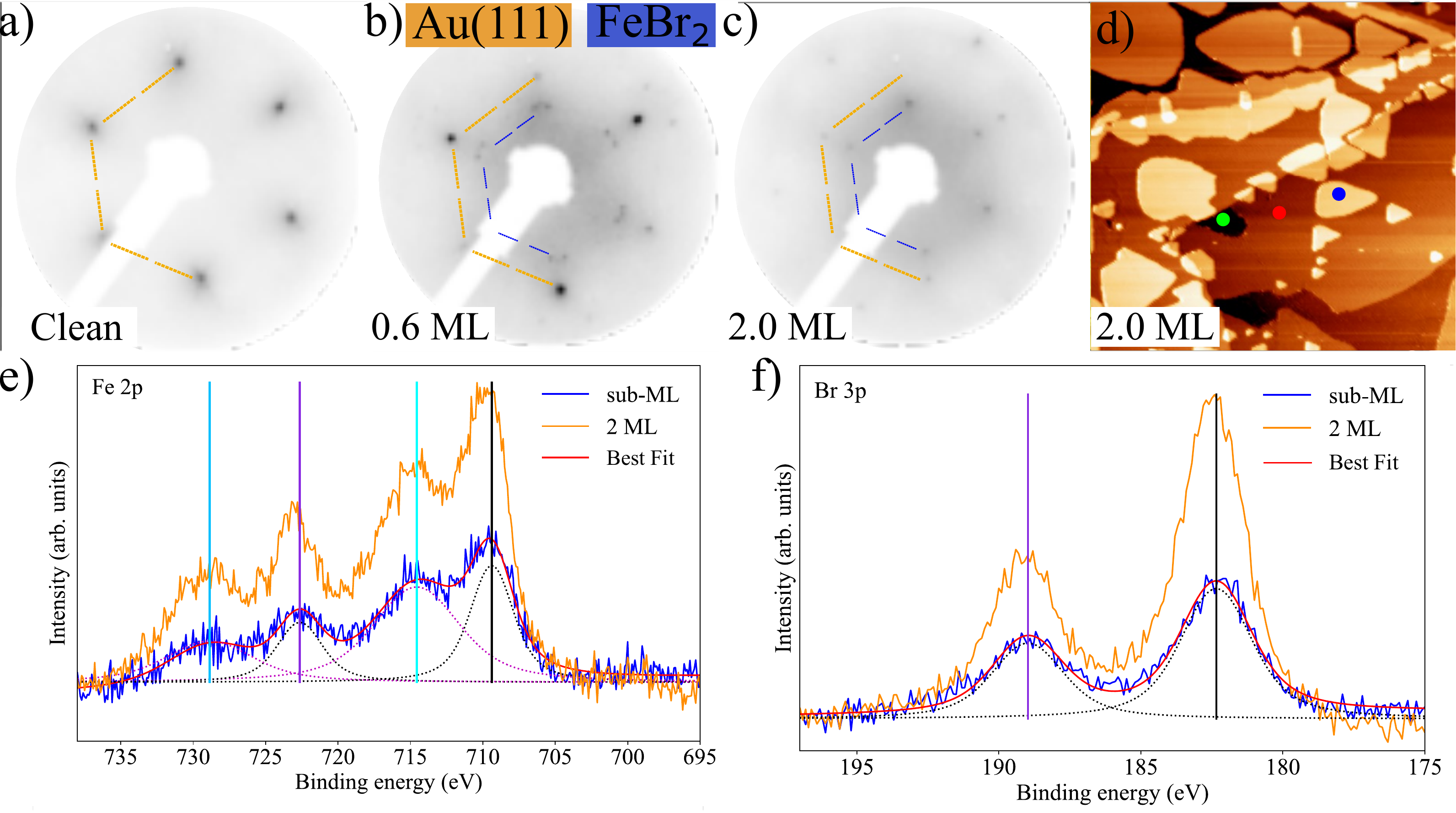}
\caption[multilayered growth]{(a-c) LEED images at 137~eV of a) clean Au(111), b) sub-monolayer (0.6-ML) and c) bilayer (2-ML) of FeBr\textsubscript{2} on Au(111). The orange half hexagon is used to designate the pattern of  Au(111) and the blue half hexagon marks the pattern from FeBr\textsubscript{2}. d) Zoomed-in part of the STM image (Fig.~\ref{fig:BOREAS_STM}) of the 2.0-ML FeBr\textsubscript{2}/Au(111), (T=77~K, U$_{\text{Bias}}=1$ V and I$_{\text{TC}}=0.02\cdot 10^{-9}$ A) measured at the BOREAS beamline. The image size is $129.9\times129.9$~nm$^{2}$. In one of the terraces we appended green, red and blue spots to indicate the levels corresponding to the first, second and third ML of FeBr\textsubscript{2}, respectively. e-f) XPS spectra of 0.6~ML and 2~ML of FeBr\textsubscript{2}/Au(111) showing the Fe 2p and Br 3p core levels.}
\label{fig:multilayer}
\end{figure*}

\begin{figure*}[ht!]
\centering
\includegraphics[width=\linewidth]{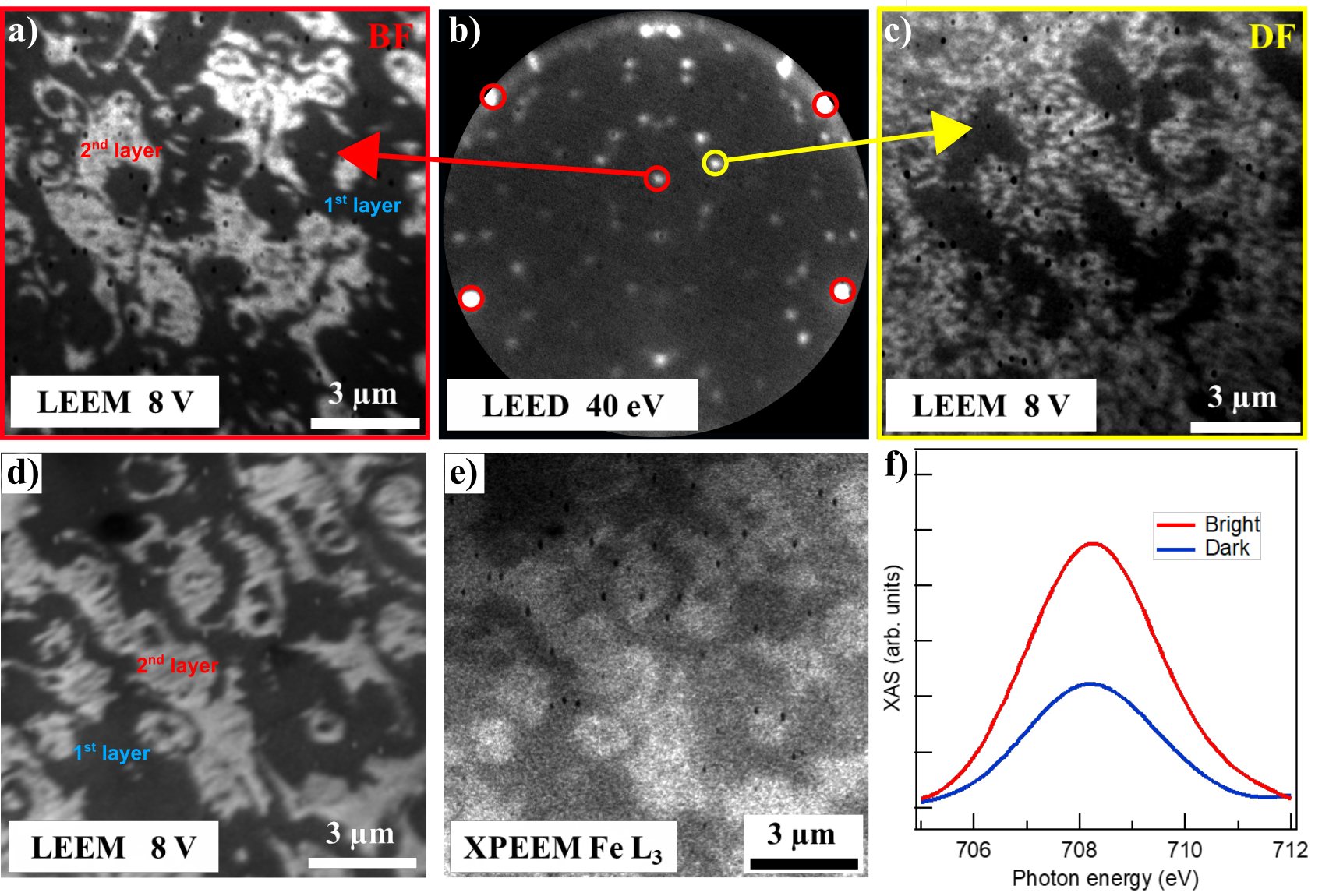}
\caption[LEEMPEEM]{LEEM and XPEEM images of 1.5 ML of FeBr\textsubscript{2} on Au(111) at room temperature. a) Bright field (BF) LEEM image. b) $\mu$-LEED pattern at 40 eV, red circles indicate the Au (111) LEED pattern and the yellow circle marks the spot, belonging to the FeBr\textsubscript{2} superstructure that was used for the dark-field (DF) image. The $\mu$-LEED pattern is distorted, since the experiment was performed with the microscope working at 10~kV, energy for which the lenses were not completely aligned in the diffraction mode, to overcome sparks during the experiment. The Au(111) pattern was used as a guide to the eye to correct the distortions. c) DF-LEEM image taken at the same area as the BF image in panel a). d) Bright-field LEEM image in a different area of the sample. e) XPEEM image at the Fe L$_{3}$-edge in the same area as panel d). f) Averaged intensities of the bright and dark areas of the XPEEM image e), as a function of the X-ray photon energy. The XAS spectra are obtained by taking the intensity of the image in certain points of the image.}
\label{fig:LEEM}
\end{figure*}

\begin{figure*}[ht!]
\includegraphics[width=\linewidth]{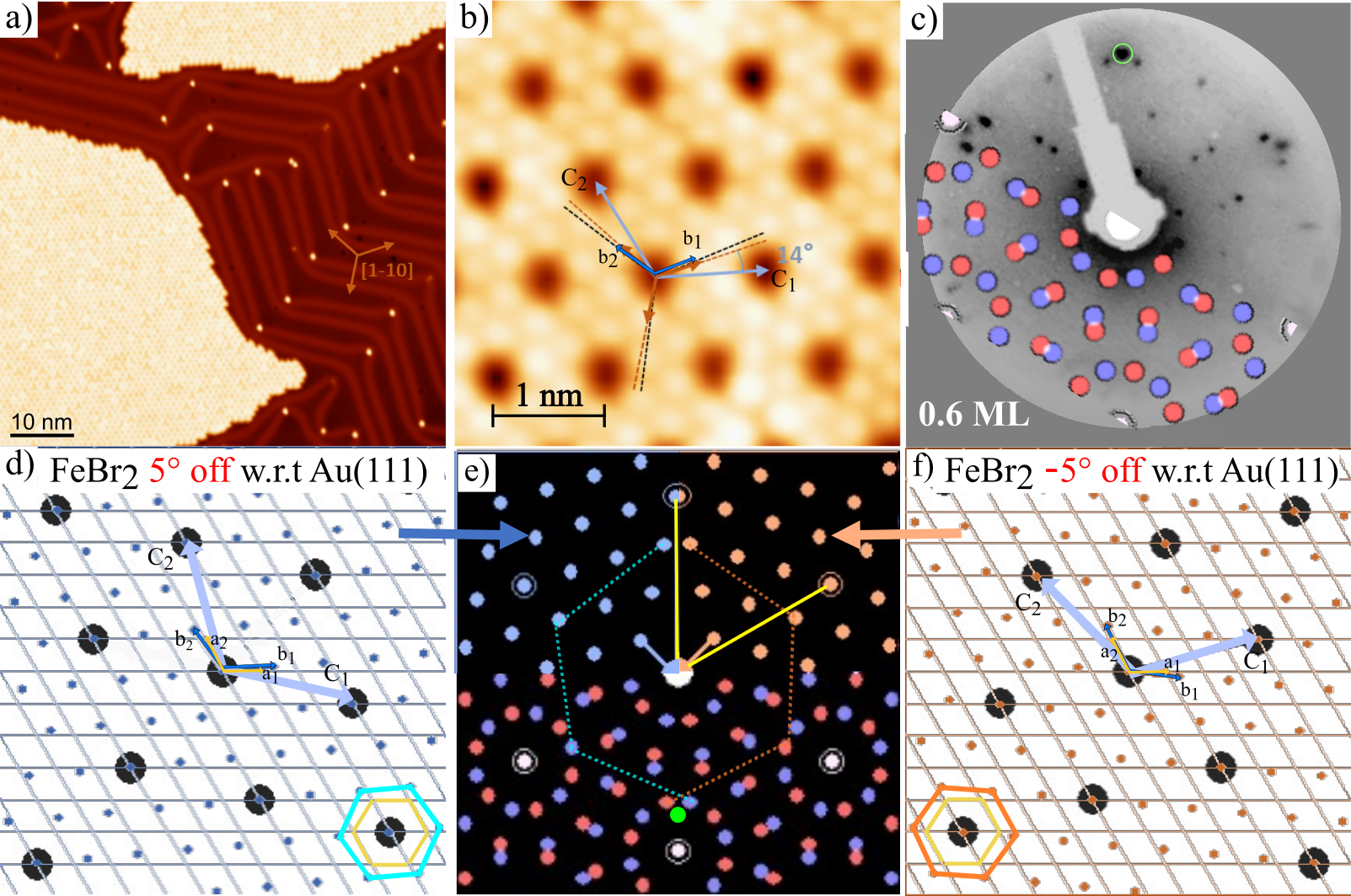}
\caption[monolayered growth]{(a-b) STM images of 0.6-ML of FeBr\textsubscript{2} at 4.3~K, measured at a) U$_{\text{Bias}}$=1~V and I$_{\text{TC}}$=0.1~nA and b) U$_{\text{Bias}}$=1 mV and I$_{\text{TC}}$=1 nA. Brown arrows indicate the close-packed Au [1$\bar{1}$0] and equivalent directions. The superstructure unit cell (light blue arrows) and the hexagonal Br lattice unit cell (black lines) are rotated with respect to the substrate high-symmetry directions. c) LEED pattern of a 0.6-ML sample measured at 43~eV, partially overlayed with the simulated pattern. The blue and red spheres are representing the two rotational domains of the superstructure. The green circle marks one of the spots belonging to the second layer FeBr\textsubscript{2} hexagonal pattern. d) Relative orientation of Au(111) and rotated Br atomic nets. The lines are representing at each crossing point the position of a Au atom, the blue dots are representing the Br atoms and the dark discs mark the coincidence points (a, b and C are used to designate the respective unit cell vectors). e) shows the simulated LEED pattern of Au(111) and of the two symmetric domains of the superstructure. Yellow lines point to the Au(111) pattern, meanwhile the dashed lines show the patterns of FeBr\textsubscript{2} for both domains. Their relative orientation represents the configurations illustrated by two hexagons in the bottom right and bottom left corners of the panels d) and f). The green dot shows the position of a spot belonging to the hexagonal pattern of the second layer of FeBr\textsubscript{2}. f) is the same as in d) but the Br grid is rotated in the opposite direction.}
\label{fig:monolayer}
\end{figure*}

\noindent The initial stage of growth of FeBr\textsubscript{2} films on single crystal Au(111) was studied by means of surface-sensitive electron diffraction and scanning tunneling microscopy. LEED patterns measured at 137~eV demonstrate a variation of the crystal structure of FeBr\textsubscript{2} with increasing number of deposited layers (Fig.~\ref{fig:multilayer} (a-c)). The hexagonal pattern characteristic of the clean Au(111) surface becomes attenuated and a new hexagonal pattern with smaller period and the same orientation appears when 0.6-ML of FeBr\textsubscript{2} is grown. An additional complex pattern of multiple dots surrounding the first-order spots of Au(111) is indicative for a surface reconstruction process that is depicted in the atomically resolved STM image (Fig.~\ref{fig:monolayer}). In the LEED pattern acquired for the 2-ML sample, the Au(111) signal is barely visible and the reconstruction-related superstructure is strongly attenuated. At this coverage  the hexagonal pattern of the ordered FeBr\textsubscript{2} becomes  a dominant motif, also seen as  second-order diffraction spots. This behaviour is characteristic of epitaxial, close to layer-by-layer growth of the overlayers on single-crystal substrates. The large-scale STM image shown in Fig.~\ref{fig:multilayer}~(d) (see also Fig.~\ref{fig:BOREAS_STM}) demonstrates that the islands of FeBr\textsubscript{2} have triangular shapes and well-defined common directions of the symmetry axes. It corroborates well with the ordered epitaxial growth inferred from the LEED diffraction patterns. One can distinguish large areas of the same thickness and limited amount of exposed atomic planes that discards a 3D growth mode. However, in the sample with nominal amount of 2.0-ML, there is nucleation of islands of the third layer and, at the same time, some voids exposing the first layer, which leads to the conclusion that the growth does not proceed in a perfect layer-by-layer mode. Fig.~\ref{fig:multilayer}~(d) shows also islands of FeBr\textsubscript{2} that grow over the atomic step of the substrate. This behaviour was reported earlier for a number of different 2D materials~\cite{Bana2018, Bets2019, Gunther2011, Yang2020}.

\noindent The chemical composition of the films was probed using XPS measurements. Survey spectra (not shown) show no traces of oxygen or other contamination. The Fe 2p and Br 3p spectra acquired for the 0.6-ML and the 2.0-ML samples as well as the calculated best-fitting curves are represented in Fig.~\ref{fig:multilayer}~(e-f). For the data evaluation, a Shirley background was subtracted and the peaks were fitted with a combination of Voigt functions (Python lmfit routine \cite{newville_lmfit_2014}). The shape of the Fe 2p spectra closely resembles the spectrum of Fe$^{2+}$ reported for thin insulating films of FeO \cite{martin-garcia_unconventional_2016,yamashita_analysis_2008} and FeCl\textsubscript{2}~\cite{Zhou_defects_2020}. In total, four Voigt profiles were needed to fit these spectra: two for the main Fe$^{2+}$ peaks and two for the satellite peaks. The main peaks of the Fe 2p core level are located at a binding energy of 709.4~eV for Fe 2p$_{3/2}$ and 722.6~eV for Fe 2p$_{1/2}$ with a spin orbit (SO) splitting of $\sim$13~eV, in close accordance with the data reported in literature \cite{martin-garcia_unconventional_2016}. The satellites are associated to the final-state effect and are related to the multiplet structure of the 2p transition metal core level \cite{Bagus_2022}. The Br 3p doublet is similar to the spectrum reported for Br$^{1-}$ in KBr and the position of the peaks falls in the same range of energies \cite{moulder_handbook_1992}. For further information about the fitting parameters see table \ref{ref:XPS}. Both Fe 2p and Br 3p spectra have the same shape for the 0.6-ML and the 2.0-ML samples. In contrast to the situation observed for NiBr\textsubscript{2}, XPS spectra of the first layer of FeBr\textsubscript{2} have no any additional components that could be interpreted as its partial decomposition~\cite{bikaljevic_noncollinear_2021}. Fitting of the Fe 2p spectra for both samples yielded the same parameters (FWHM and center position). Variation of the FWHM for the Br 3p peaks (see table \ref{ref:XPS}) can be attributed to the different environment of the bottom Br layer interfaced with the Au(111) surface and higher Br layers~\cite{bana_epitaxial_2018}. The peak ratio between the main and the satellite peaks as well as the calculated ratio for Br and Fe stays constant for both samples, supporting the presence of one single stoichiometric phase of FeBr\textsubscript{2} in the ordered layers epitaxially grown on Au(111).  

\noindent The uniform epitaxial growth was also verified on the mesoscopic scale via LEEM and XPEEM measurements, performed at room temperature. We use the capability of LEEM microscopy and XPEEM to provide information of images with structural and chemical contrast~\cite{Flege2014}, to study the 1.5-ML sample that was grown in-situ in the preparation chamber of the microscope. The bright-field LEEM image (the image obtained with a specular-00 spot) is shown in Fig.~\ref{fig:LEEM}~(a). The contrast arises from the difference in the local reflectivity of the film with variable thickness~\cite{Flege2014}. The image represents one complete layer and large, $\mu$m-scale islands of the second layer in close accordance with the results of the STM measurements (Fig. \ref{fig:multilayer} (d)) and \ref{fig:BOREAS_STM}). For identification of the layers we make use of the reconstruction, characteristic of the first layer of FeBr\textsubscript{2}/Au(111). The pattern acquired for the 1.5-ML sample with 40~eV energy of electrons (Fig. \ref{fig:LEEM} (b)) is a superposition of the complex LEED pattern Fig.~\ref{fig:multilayer}~(b) and the hexagonal pattern Fig.~\ref{fig:multilayer}~(c), originated from the second layer of FeBr\textsubscript{2}. Presence of the first-order Au(111) spots shows that the first layer of FeBr\textsubscript{2} is not perfectly continuous. Selecting a diffracted beam from the FeBr\textsubscript{2} superstructure, i.e, centering the illumination deflectors at that LEED spot, a dark-field image in real space was formed (Fig. \ref{fig:LEEM} (c)). It has inverted contrast with respect to the BF image Fig. \ref{fig:LEEM} (a). The areas with higher intensity are those that feature the reconstruction's LEED pattern (the first layer of FeBr\textsubscript{2}). Variation of contrast within the bright zones occurs because of the presence of two rotational domains inside of the first layer of FeBr\textsubscript{2} (see discussion related to the Fig.~\ref{fig:monolayer}). The large dark areas display no reconstruction. To prove that these are indeed the areas occupied with 2~ML  of FeBr\textsubscript{2} but not the pure Au(111), BF LEEM (Fig.~\ref{fig:LEEM}~(d)) and XPEEM (Fig.~\ref{fig:LEEM}~(e)) images were acquired at the same position.

\noindent Again, the specular (00)-spot of the LEED pattern was used for the LEEM measurements, therefore the contrast in Fig.~\ref{fig:LEEM}~(a) and \ref{fig:LEEM}~(d) is the same. The XPEEM image shows the local difference in X-ray absorption at the Fe L$_3$-edge. The averaged intensities of the bright and dark zones were calculated and represented in Fig. \ref{fig:LEEM}~(f) as a function of the X-ray photon energy. A larger absorption peak characteristic of the bright zones in the XPEEM image proves a higher thickness of the FeBr\textsubscript{2} in these areas and consequently in the bright areas of the BF LEEM images (Fig.~\ref{fig:LEEM}~(a) and \ref{fig:LEEM}~(d)). Combining these results with our previous observations, we can conclude that the results of the LEEM/XPEEM characterization corroborate that the growth of FeBr\textsubscript{2} on Au(111) is close to the layer-by-layer mode.

\noindent Investigation of the atomic arrangement that gives rise to the reconstruction of the first layer of FeBr\textsubscript{2}/Au(111) was performed using LT-STM and LEED. The STM image displayed in Fig. \ref{fig:monolayer}~(a) shows two islands of FeBr\textsubscript{2} separated by the bare Au(111) surface with the characteristic $22\times\sqrt{3}$ herringbone reconstruction~\cite{murphy_impact_2015,schouteden_fourier-transform_2009}. Apart of bright dots at the elbows of the herringbone probably associated with initial nucleation of FeBr\textsubscript{2}, the Au(111) remains clean and the FeBr\textsubscript{2} grows as compact ordered islands. A zoom-in image in Fig. \ref{fig:monolayer} (b) reveals details of a superstructure in the first layer of FeBr\textsubscript{2}/Au(111) with atomic-resolution. It consists of a triangular net of dark spots with periodicity of $9.7 \pm 0.72\text{ \AA}$ that obscure single Br atoms in otherwise flat layer.

Interatomic distances in the top-most Br layer were found to be of $3.66\pm 0.3\text{ \AA}$, in reasonable agreement with the expected monolayer lattice constant calculated by DFT \cite{botana_electronic_2019} and the bulk value of $3.78\text{ \AA}$ for FeBr\textsubscript{2}~\cite{haberecht_refinement_2001} (see also Fig. \ref{fig:InteratomicLines}). The angle between the closed-packed directions of Au(111) and of the top-most Br plane is $\sim 5^{\circ}$, meanwhile the angle between the high-symmetry directions of the Au(111) and the lattice vectors of the superstructure is $\sim 14^{\circ}$. Fig. \ref{fig:monolayer} (b) shows that the unit vectors c\textsubscript{1} and c\textsubscript{2} of the reconstruction can be represented in terms of the unit vectors b\textsubscript{1} and b\textsubscript{2} of the Br plane as: 
\begin{equation}\label{eq:m1}
\begin{pmatrix}
c_{1} \\
c_{2}
\end{pmatrix}=
\begin{pmatrix}
2 & -1 \\
1 & 3
\end{pmatrix}\cdot
\begin{pmatrix}
b_{1} \\
b_{2}
\end{pmatrix},
\end{equation}
where we drop the vector sign. Although the top and the bottom Br planes in the 1T structure of a single FeBr\textsubscript{2} slab are not equivalent, they have the same orientation of the high-symmetry directions and their lateral positions can be obtained by a rigid  shift along the b\textsubscript{1}-b\textsubscript{2} direction. For the sake of clarity, we do not distinguish the top from the bottom Br planes considering relative orientation of the Br and Au(111) layers, keeping in mind this relative shift.

\noindent In Fig. \ref{fig:monolayer}~(d), the Au(111) plane is represented by two series of equally spaced parallel lines, crossed at $120^{\circ}$, and the Br plane is displayed as the set of ordered dots with six-fold symmetry. The angle between the close-packed directions of these layers is set to~$5^{\circ}$. It is clearly seen that the  vectors c\textsubscript{1} and c\textsubscript{2} constructed in accordance with Eq.~(\ref{eq:m1}) point to the places of coincidence between the Au(111) and the Br layers. Using the unit vectors a\textsubscript{1} and a\textsubscript{2} of the Au(111) plane, they can be represented in a matrix form as: 
\begin{equation}\label{eq:m2}
\begin{pmatrix}
c_{1} \\
c_{2}
\end{pmatrix}=
\begin{pmatrix}
3 & -1 \\
1 & 4
\end{pmatrix}\cdot
\begin{pmatrix}
a_{1} \\
a_{2}
\end{pmatrix}.
\end{equation}
These points and the equivalent ones are marked with large dark discs in the Fig. \ref{fig:monolayer}~(d). Calculations presented in the appendix shows that exact coincidence requires rotation of the Br layer by $5.21^{\circ}$ and lateral expansion of the FeBr\textsubscript{2} by~$\sim 3\%$ with respect to the bulk value. 

\noindent The symmetry of the system requires the existence of FeBr\textsubscript{2} islands rotated by the same angle with respect to the Au(111) but in the opposite direction. This situation is shown in Fig. \ref{fig:monolayer}~(f). Representation of the unit vectors $\Tilde{c}_1$ and $\Tilde{c}_2$ of the coincidence points in terms of the unit vectors of the Br plane $\Tilde{b}_1$ and $\Tilde{b}_2$ as well as the unit vectors of the Au(111) a\textsubscript{1} and a\textsubscript{2} look like:     
 \begin{equation}\label{eq:m3}
\begin{pmatrix}
\Tilde{c}_{1} \\
\Tilde{c}_{2}
\end{pmatrix}=
\begin{pmatrix}
3 & 1 \\
-1 & 2
\end{pmatrix}\cdot
\begin{pmatrix}
\Tilde{b}_{1} \\
\Tilde{b}_{2}
\end{pmatrix}=
\begin{pmatrix}
4 & 1 \\
-1 & 3
\end{pmatrix}\cdot
\begin{pmatrix}
a_{1} \\
a_{2}
\end{pmatrix}.
\end{equation} 

\noindent Fig. \ref{fig:monolayer}~(e) shows a simulation of the LEED pattern by means of the LEEDpat software~\cite{hermann_leedpat_nodate}. We used Au(111) as a substrate and the superstructure visible in the STM image (Fig. \ref{fig:monolayer}~(b)) as the dark dots was represented by an artificial overlayer. The Au(111) unit-cell size was taken to be $2.86 \text{ \AA}$. The overlayers were defined using the matrix relations Eq.~\ref{eq:m2} and Eq.~\ref{eq:m3}, respectively. Results of the simulation for each domain are shown in the top-left and top-right quarters of Fig. \ref{fig:monolayer}~(e), while the bottom half of the figure shows a superposition of both patterns. We can clearly see the characteristic twelve-point circles around the central and the first-order Au(111) spots observed in Fig. \ref{fig:multilayer}~(b) and \ref{fig:LEEM}~(b). A 43-eV LEED pattern taken for the 0.6-ML FeBr\textsubscript{2}/Au(111) sample is shown in Fig. \ref{fig:monolayer}~(c). This is the same sample that was used to obtain the LEED data in Fig. \ref{fig:multilayer}~(b), but the superstructure pattern is much more pronounced here because of the lower energy of electrons. Half of the image is overlayed by the simulated pattern of the superstructure. An additional point highlighted with a light green circle belongs to the hexagonal pattern of the second layer of FeBr\textsubscript{2} (corresponding patterns in Fig. \ref{fig:multilayer}~(b) and (c) are marked with a blue hexagon). Since the domains of the reconstructed first layer of FeBr\textsubscript{2} are rotated by ~$\pm 5^{\circ}$ with respect to the gold lattice, but the second layer grows aligned with the substrate, \and since atomically resolved STM images of the second and third layers of FeBr\textsubscript{2} shown in Fig.~\ref{fig:BLTL} display no signs of the reconstruction  of the underlying layer, we conclude that the superstructure disappears upon the growth of the higher layers.

\noindent The dark spots observed in the STM image (Fig. \ref{fig:monolayer}~(b)) either represent some sort of defects that follow the period of the coincidence points between the Au(111) lattice and the Br lattice of the first FeBr\textsubscript{2} layer or can be a pure electronic effect arising due to the interaction between the Br and Au atoms at the interface.  Defects in the isostructural compound FeCl\textsubscript{2}, which were simulated~\cite{ceyhan_electronic_2021} and studied experimentally~\cite{Zhou_defects_2020}, have different appearance. 
We have observed similar objects as in Ref. ~\cite{Zhou_defects_2020} randomly distributed within the first layer of FeBr\textsubscript{2}/Au(111) (see Fig.~\ref{fig:InteratomicLines} and \ref{fig:STS}~(b)). Although we were unable to measure the bandgap, our STS data (Fig. \ref{fig:STS}~(a)) show a bump situated at 0.4~eV with respect to the Fermi level that we interpret as the onset of the conduction band (CB), meanwhile the valence band is below the range of our measurements (-2~eV). Therefore, we suggest that a monolayer of FeBr\textsubscript{2}/Au(111) is an insulator. In this case, screening of the charge imbalance that would be the consequence of an atomic vacancy would be impeded and such a defect would affect the electronic state of the surrounding atoms. Indeed, this effect was visualised in the STM image (Fig. \ref{fig:STS}~(b)) and the corresponding conductance map, displayed in Fig. \ref{fig:STS}~(c). Furthermore, if these spots were defects, we would expect some random imperfections in their ordered structure unavoidable in any real system, which were never seen in our STM data. On the other hand, STM images of the reconstructed first layer of FeBr\textsubscript{2}/Au(111) obtained at different values of the bias voltage do not show any change of contrast that we would expect if the dark dots of reconstruction were a purely electronic effect (see figure~\ref{fig:BiasDep}). Therefore, the mechanism leading to the reconstruction of the first layer of FeBr\textsubscript{2}/Au(111) is not yet defined and requires further investigation.

\subsection{Magnetic properties}

\begin{figure}[h!]
\centering
\includegraphics[width=\linewidth]{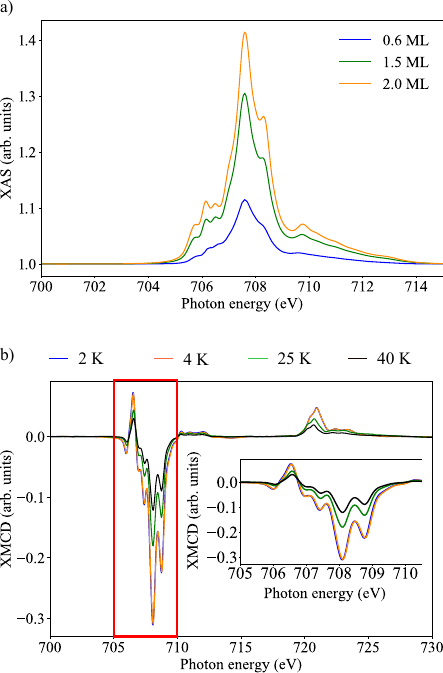}
\caption[XAS and XMCD different thickness and temperature]{a) White-line XAS spectra for different thicknesses measured at the Fe L\textsubscript{3}-edge, 6~T, 2~K and NI. The spectra are shifted along the energy axis to align the position of the  Fe L\textsubscript{3} peak maxima. Further information about the shift corrections is available in the supplementary information, Fig. \ref{fig:ShiftCorrect}. The background subtraction was performed by using asymmetrically reweighted penalized least squares smoothing \cite{baek_baseline_2015}. b) XMCD spectra of 2.0-ML FeBr\textsubscript{2} on Au(111) measured at 6~T and NI for different temperatures. The inserted image is a zoomed-in version of the L$_{3}$ region. All measurements displayed in a) and b) were performed at the BOREAS beamline.}
\label{fig:xasxmcd}
\end{figure}

\noindent Magnetic properties of the in-situ-grown single-and double-slab films were measured via XAS/XMCD using circularly polarized synchrotron X-ray radiation. White line (average of the spectra with left and right polarisation) absorption spectra at the Fe L\textsubscript{3} edge aligned to the maximum of the peak and the respective XMCD spectra are shown in Fig.~\ref{fig:xasxmcd}. The structure of the XAS peak closely resembles the Fe L\textsubscript{3} XAS spectrum measured for FeCl\textsubscript{2} \cite{everett_ferrous_2014}, which was attributed to the Fe$^{2+}$ oxidation state~\cite{kowalska_iron_2017,miedema_iron_2013} (see also Fig.~\ref{fig:Fe2+}). It does not vary neither with thickness nor with temperature, which confirms the observation from the analysis of the XPS data that the films are uniform, single-phase and contain Fe$^{2+}$ ions in the same coordination.

\begin{figure*}[ht!]
\centering
\includegraphics[width=\linewidth]{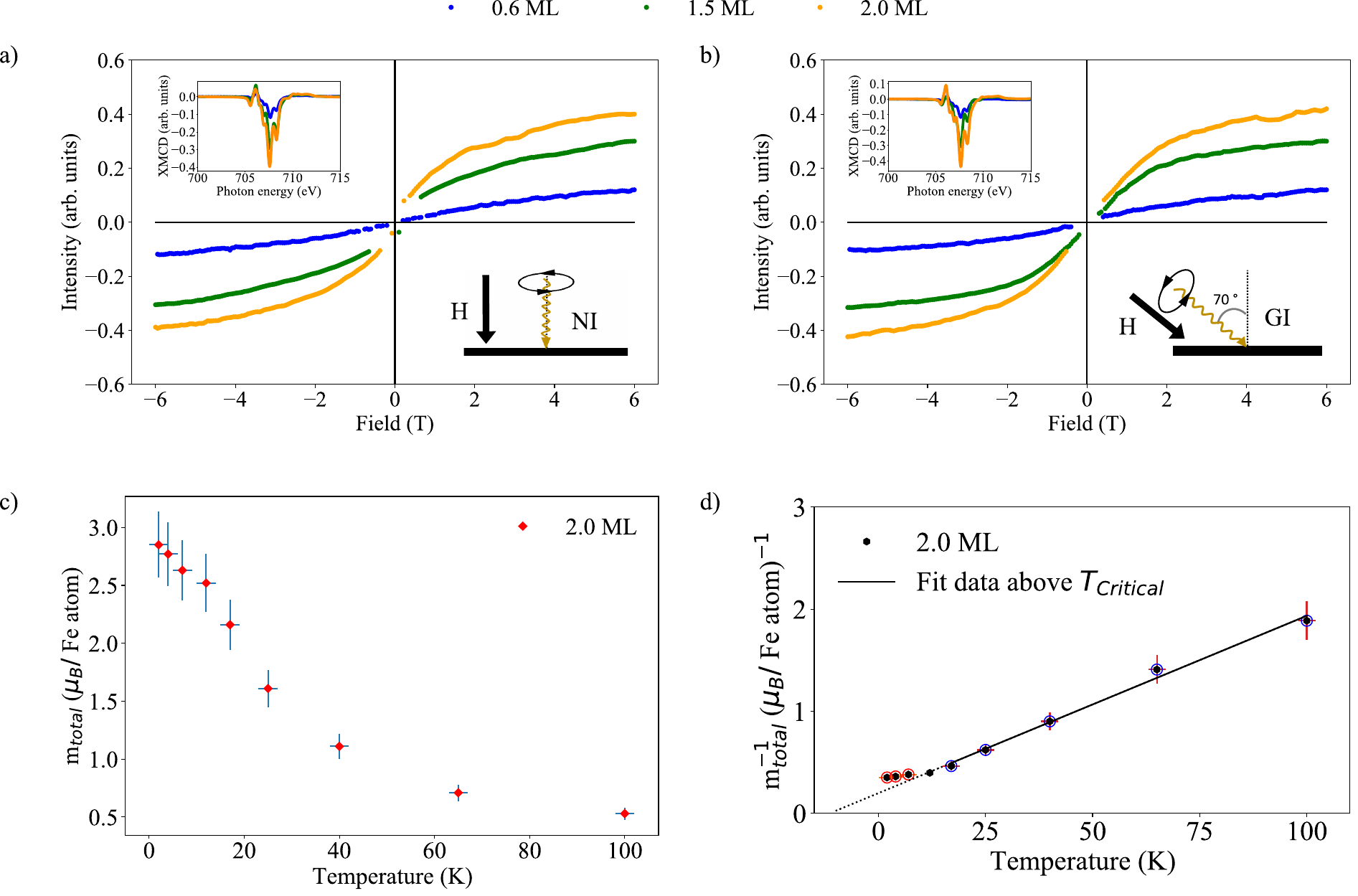}
\caption[Magnetization curve different sample thickness]{Comparison of the XMCD magnetization loops for 0.6-ML, 1.5-ML and 2.0-ML FeBr\textsubscript{2}/Au(111) films measured at 2~K at: a) normal incidence (NI) and b) grazing incidence (GI) at the Fe~L\textsubscript{3}-edge, with the field ($H$) parallel to the beam direction. The magnetization curves are normalized to the respective white line (averaged ($\sigma^{+}$+$\sigma^{-}$)/2) XAS peak height. The insert shows the corresponding Fe~L\textsubscript{3} XMCD peaks at 6~T and 2~K normalized to the isotropic XAS edge jump for all three samples. The magnetization loops are normalized to the maximum intensity at 6~T and multiplied by the XAS edge-jump height of the L\textsubscript{3}-edge.
Calculated effective spin moment from the spectra of the 2.0-ML sample at 6~T and NI (c) and it's inverse (d) for different temperatures. The hollow blue circles are representing the data, which were used for the linear fit of the high temperature regime (17-100~K$>T_{\text{Critical}}$). The excluded data points are displayed in red. A slope of $0.017 \pm 0.001$ $(K \cdot \mu_{B}/\text{Fe atom})^{-1}$ and a linear intercept of $-0.198 \pm 0.049$ $(\mu_{B}/\text{Fe atom})^{-1}$ ($\chi^{2}=0.01$) gives the estimation of the paramagnetic Curie temperature of $-10\pm 1$~K. The chosen temperature range for the performed fit was based on the fact that the lowest used temperature is far away from saturation of a Brillouin function in a J=2 system. At 17~K x would be 1.9 (not in saturation see Fig. \ref{fig:BrillX}). All measurements displayed in a) and b) were performed at the BOREAS beamline.}
\label{fig:hyst}
\end{figure*}

\noindent XMCD magnetization curves measured for different thicknesses at 2~K in two different geometries: normal incidence (NI) and grazing incidence (GI), are shown in Fig.~\ref{fig:hyst} (a-b). Since they were measured in total electron yield (TEY) mode, the curves have artifact spikes around 0~T, which were removed. The loops are normalized to the Fe L\textsubscript{3} peak height of the respective white line (2~K, NI or GI) spectra and therefore the intensity values are proportional to the projection of the thermal average of the magnetic moment per Fe atom on the x-ray beam direction at 6 T field. The corresponding XMCD spectra are displayed as the inset in Fig.~\ref{fig:hyst} (a-b). The loops do not show any field hysteresis outside the range affected by TEY artifacts and magnetization vanishes close to zero field, implying that the samples possess no simple collinear ferromagnetic ordering at that temperature.\\

\begin{table}[ht!]
\begin{tabular}{c|c|ccccc|}
\hline
\multirow{3}{*}{ML} & \multirow{3}{*}{T (K)} & \multicolumn{4}{c}{$\mu$ ($\mu_B$/Fe atom)}                                                                                                             \\ \cline{3-6} 
                               & \multicolumn{1}{c|}{}                      &                        & \multicolumn{1}{c|}{NI}                                                      & \multicolumn{1}{c}{GI}                                                      \\ \cline{3-6} 
                                 &   & \multicolumn{1}{l}{$m_{\text{Seff}}$} & \multicolumn{1}{l}{$m_l$} & \multicolumn{1}{l}{$m_{\text{Seff}}$} & \multicolumn{1}{l}{$m_l$} \\ \hline
0.6                            & 2                      & 1.13                   & 0.30                     & 1.05                   & 0.36                                         \\ \hline
1.5                                                                & 2                      & 1.814                  & 0.60                     & 2.03                   & 0.45                                        \\ \hline
2.0                             & 2                      & 2.15                   & 0.70                     & 1.91                   & 0.47                                        \\ \hline
\end{tabular}
\caption[Magnetic moment for different coverages lowest temperature]{Magnetic moments calculated by means of the sum rules from XMCD spectra obtained at T=2~K and B=6~T. Magnetic moments ($\mu$) are divided in two sections for NI and GI. The error for each magnetic moment is $\pm 10\%$. More details about the procedure of sum-rule analysis and the extended version of the data (Table~\ref{Tab:MagMom}) are available in the supplementary information.} 
\label{tab:sum_rules_brief}
\end{table}
               
\noindent It was demonstrated in the previous section that growth of FeBr\textsubscript{2} is close to the layer-by-layer mode. Therefore the 0.6-ML sample comprises mainly one layer thick islands while the 1.5-ML and 2.0-ML  samples consist of the complete first slab and islands of the second and, in minor proportion, of the third slab (see also the STM images of the 1.5-ML and 2.0-ML samples acquired at the beamline, Fig.~\ref{fig:BOREAS_STM}). It is clearly seen from the loops in Fig.~\ref{fig:hyst} that the expectation value of the magnetization at 2~K and 6~T along the beam direction is substantially lower in the first layer than in the thicker films. Different sub-ML samples were grown directly at the beamline and compared to the sample transferred via suitcase. Therefore, we can exclude that the reduced magnetization is a result of contamination, since all samples, also the ones grown directly at the beamline, showed a strongly reduced magnetization. Sum-rule analysis of the spectra yields values of the spin magnetic moment close to 1~$\mu_B$ for the 0.6-ML sample and about 2~$\mu_B$ for the 1.5-ML and 2.0-ML samples (see Table~\ref{tab:sum_rules_brief}). 

\noindent Analysis of a heterogeneous magnetic system is complicated, nevertheless we can infer some additional information about magnetic ordering in the thin films of FeBr\textsubscript{2} from the temperature dependence of their magnetic properties. Fig.~\ref{fig:hyst} shows in panels (c) and (d) the expectation value of the spin magnetic moment of the 2.0-ML sample and its inverse as a function of temperature (respective values are listed in Table~\ref{Tab:MagMom}), as obtained by sum-rule analysis of the spectra at 6~T. In the paramagnetic state, $1/m_{total}$  is proportional to the susceptibility measured in the field of 6~T, provided that the temperature is high enough to yield a linear magnetization curve. In Fig.~\ref{fig:BrillX}, we show that for Brillouin functions with J = 2 and T$\geq$ 17 K, this is approximately the case. Linear fitting of the high-temperature part of this curve yields an extrapolated ordering temperature of $\sim -10$~K, well below the value of 3.5~K reported for the bulk material~\cite{mcguire_crystal_2017}. Its negative sign indicates the possibility of antiferromagnetic behaviour. Although a non-zero value of the extrapolated ordering temperature cannot serve as unambiguous prove, it implies that the 2.0-ML sample can be magnetically ordered at 2~K. Panel (a) of Fig.~\ref{fig:BrillouinFit} shows the same total magnetic moments of the 2.0-ML sample together with curves calculated using the Brillouin function. We observe a slope change of the $m_{total}$ vs T curve at $\sim 7$~K, in contrast to the behavior expected from a paramagnetic system with S = 2, supporting the presence of magnetic correlations in the 2.0-ML sample.

\noindent The magnetic behaviour of the single-layer FeBr\textsubscript{2} is distinctly different. Low values of the magnetic moments in an external field of 6~T as compared to the thicker films (Table~\ref{tab:sum_rules_brief}) and shallow magnetization loops in both NI and GI directions (Fig.~\ref{fig:hyst}) imply a magnetic order that is neither paramagnetic nor ferromagnetic. Figure \ref{fig:BrillouinFit} demonstrates that fitting of the NI 0.6-ML sample's loop to the Brillouin function does not yield satisfactory results. The paramagnetic curve at 2~K is steeper than the experimental loop and has a smaller tangent in the high-field regions. Including out-of-plane anisotropy in the model of the paramagnetic system would make it's loop even steeper. Therefore, a mere lack of magnetic order cannot explain the observed magnetic properties of a single layer of FeBr\textsubscript{2}. STM data show that the 0.6-ML FeBr\textsubscript{2}/Au(111) sample comprises of single layer islands with lateral size of $\sim 100$~nm (see Fig.~\ref{fig:monolayer}). These islands are large enough to neglect the effect of thermal excitations at 2~K and to discard superparamagnetic behaviour. Although recent neutron diffraction data unveiled some clues of a non-collinear magnetic order in bulk FeBr\textsubscript{2}~\cite{katsumata_neutron-scattering_1997, Binek2000}, neither these nor older works~\cite{wilkinson_neutron_1959} showed antiferromagnetic order within the layers of FeBr\textsubscript{2}. At the same time, the intralayer exchange coupling to the nearest neighbour $J_1$ was found to be of different  sign with respect to the next-nearest neighbour exchange coupling constant $J_2$~\cite{katsumata_neutron-scattering_1997}. These competing interactions lead to frustration, which, according to theoretical calculations~\cite{Leonov2015}, can result in complex magnetic textures. The phase diagram presented in \cite{Leonov2015} demonstrates that the magnetic structure varies with the change of the $J_1$/$J_2$ ratio or due to modification of the anisotropy. Since the superstructure observed in the first layer of FeBr\textsubscript{2}/Au(111) causes lateral expansion of the FeBr\textsubscript{2} crystal lattice by~$\sim 3\%$ (according to our model) and the superexchange interaction depends strongly on the angle between the Fe-Br-Fe bonds, variation of the $J_1$/$J_2$ ratio can be sufficient to stabilize one of the magnetic textures predicted in~\cite{Leonov2015}. 
On the other hand, more complex situations including a spin-glass phase or complex antiferromagnetic ordering cannot be ruled out. Although, among different possible reasons for the distinctive magnetic behaviour observed in the single-layer of FeBr\textsubscript{2}, formation of non-collinear magnetic texture due to frustration looks rather probable, further investigations are necessary to unambiguously establish the type of magnetic structure.

\subsection{Summary}
\noindent Thin films (sub-ML to 2.0-ML) of FeBr\textsubscript{2} were grown epitaxially on a single-crystal Au(111) substrate in UHV via  sublimation of the stoichiometric powder compound from a Knudsen cell. Thorough characterization performed by means of XPS and XAS/XMCD spectroscopy, as well as via surface-sensitive LEED and LT-STM, LEEM and XPEEM shows that FeBr\textsubscript{2} maintains its stoichiometric chemical composition down to the single-layer limit. The growth of the films is close to the layer-by-layer mode. The first layer of FeBr\textsubscript{2}/Au(111) demonstrates an atomic reconstruction that is interpreted as the coincidence pattern between the lattice of Br atoms in $\pm 5^{\circ}$ rotated domains of $\sim 3\%$ laterally expanded FeBr\textsubscript{2} and the Au(111) surface lattice.

\noindent XMCD measurements reveal thickness-dependent magnetic properties of the FeBr\textsubscript{2}. While the temperature behaviour of the saturation magnetization of the second and higher layers shows some clues of magnetic ordering, magnetic properties of the single layer of FeBr\textsubscript{2} were found to be distinctly different. Shallow magnetization loops at 2~K, lack of saturation and low magnetization in fields up to 6~T leaves a possibility for various interpretations starting from magnetic frustration, characteristic of the triangular lattice of the magnetic Fe atoms to more complex types of magnetic order or spin-glass behaviour. These findings open the prospect for further investigation of the monolayers of the 2D magnetic transition metal dihalides. In contrast to the trihalides family that features the honeycomb arrangement of the magnetic atoms within the 2D layers, triangular nets of magnetic atoms in TMDH are prone to frustration that leads to degeneracy of the magnetic ground state and potentially may result in stronger response towards external stimuli. This quality might result in a rich variety of interesting physical phenomena and opens a way for using 2D magnetic TMDH compounds in applications.     

\section*{Acknowledgment}
\begin{acknowledgments}
\noindent C.G.-O. and M.P.-D. acknowledge funding of the Ph.D. fellowship from the MPC Foundation.
\\
S.E.H. thanks the whole AG Kuch and in particular J. Gördes for help during the BESSY measurements. Also he is very thankful to the local IT /electronics workshop/fine mechanics workshop team for their continuous support. In particular he is very thankful for the possibility to perform some last STM measurements at the PEARL beamline at SLS thanks to Dr. Matthias Muntwiler.
\\
J.N. thanks the Deutsche Forschungsgemeinschaft (DFG, German Research Foundation) for his funding under the project 277101999 - CRC 183.
\\
S.T. acknowledges financial support by the BMBF through project VEKMAG (BMBF 05K19KEA).
\\
P.G. acknowledges funding from PID2020-116181RB-C32 and FlagEraSOgraphMEM PCI2019-111908-2 (AEI/FEDER)
\\
D. G. O. acknowledges funding by the Spanish MCIN/AEI/ 10.13039/501100011033 and by the European Union "NextGenerationEU"/PRTR (PID2019- 107338RB-C63 and TED2021-132388B-C43).
\\
C. R., M. I., C.G.-O. and M.P.-D. acknowledge funding by the European Union’s Horizon 2020 research and innovation programme (grant agreement No 800923), the Spanish MCIN/AEI/ 10.13039/501100011033 (PID2020-114252GB-I00, PID2019-107338RB-C63, TED2021-130292B-C42), Basque Goverment IT1591-22,  and by the IKUR Strategy under the collaboration agreement between Ikerbasque Foundation and MPC on behalf of the Department of Education of the Basque Government.
\\
C. R., M. I., C.G.-O. and S.E.H. are very thankful for the help during the XMCD and STM measurements of Samuel Kerschbaumer, Andrea Aguirre Baños and Amitayush Jha Thakur.
\end{acknowledgments} 

\newpage

\section{Supplementary Information}
\beginsupplement
\subsection{FeBr\textsubscript{2} crystal structure}
\begin{figure}[h!]
\centering
\includegraphics[width=\linewidth]{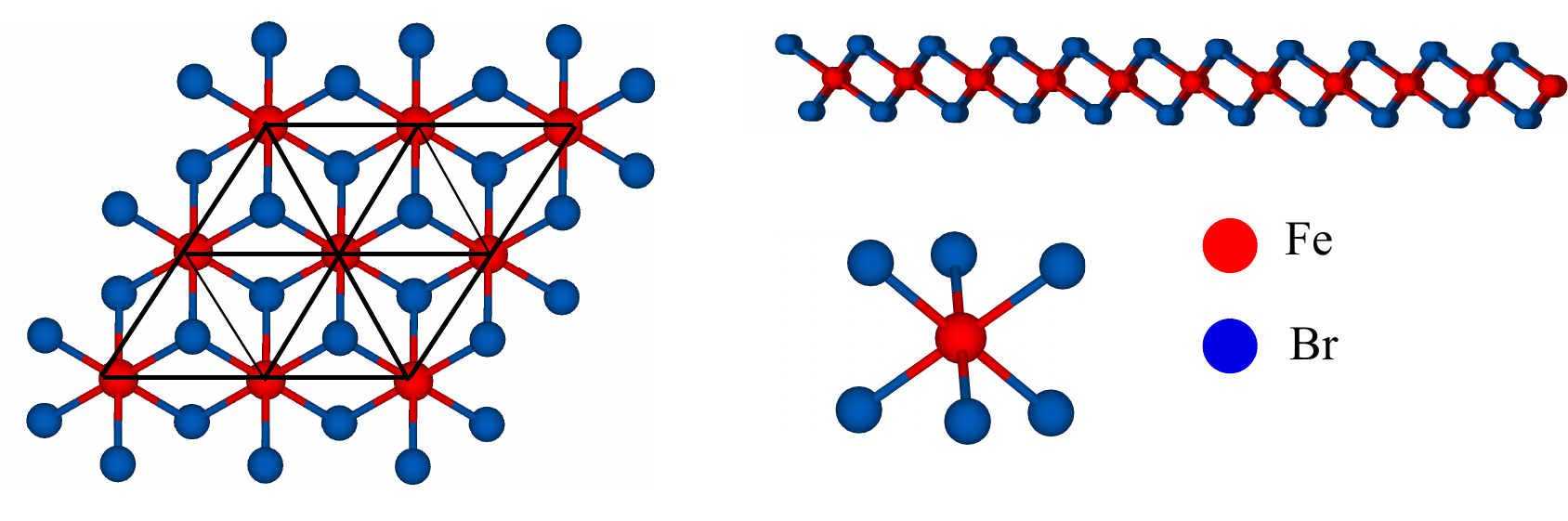}
\caption[Crystal structure]{On the left the top- and on the right the side-view of FeBr\textsubscript{2} is displayed. The blue spheres are representing the Br atoms and red spheres the Fe atoms. Image based on \cite{available_materials_2020,momma_vesta_2011}.}
\label{fig:TopView}
\end{figure}

\noindent In Fig. \ref{fig:TopView} the structure of the octahedral FeBr\textsubscript{2} is displayed from the top and from the side. The system has a lattice constant of $a=3.776 \text{ \AA}=b$ and a c-lattice constant of $6.558 \text{ \AA}$. The red spheres are representing the Fe atoms and the blue spheres the Br-atoms. On the lower right side the 1T-structure of FeBr\textsubscript{2} is displayed with the typical 180$^{\circ}$ rotated Br-planes.
The visualization of the crystal structure of FeBr\textsubscript{2} was done via VESTA \cite{momma_vesta_2011}.

\subsection{XAS based thickness calibration} 
\noindent To calculate the thickness of the samples, we first need to calculate the areal density of Fe in FeBr\textsubscript{2}.
The thickness calibration is then based on Ref. \cite{gopakumar_spin-crossover_2013}. 

\noindent Therefore, we are using the BL LEED image and construct a triangle between the (0,0)-spot and two other spots of the FeBr\textsubscript{2} hexagon. We calculate the area of the triangle by using the bulk FeBr\textsubscript{2} lattice constant of $3.776\text{ \AA}$. Here we use the fact that the Br-Br distance is the same as Fe-Fe distance in case of a 1T-structure \cite{mcguire_crystal_2017}.
The number of Fe atoms inside the triangle is $0.5$ (each corner of the triangle consists of 1/6 of an Fe atom).

\begin{figure}[h!]
\centering
\includegraphics[width=\linewidth]{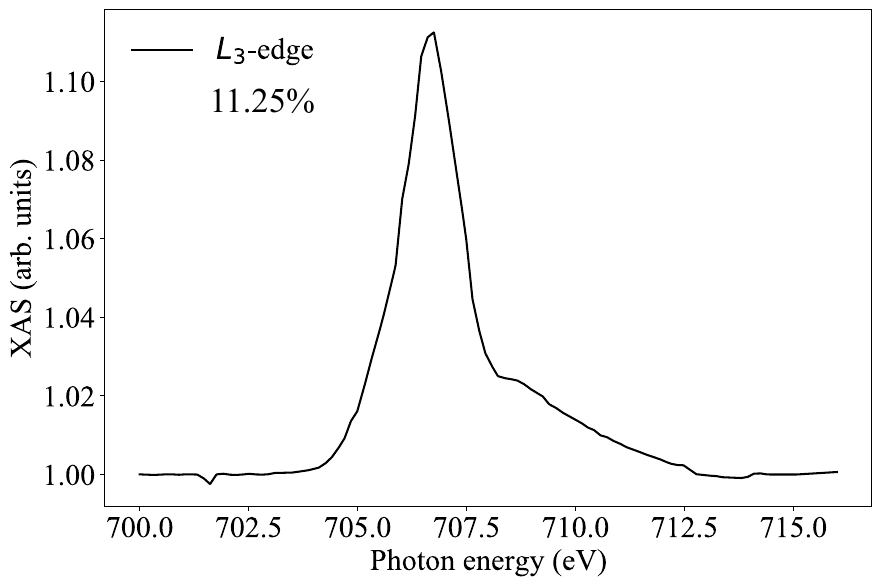}
\caption[Linear polarized Fe L$_{3}$ edge]{XAS spectra of the sub-ML Fe L$_{3}$ edge at RT without any applied field by using linearly polarized light under an angle of 55$^{\circ}$ (magic angle). This sample was measured at the VEKMAG beamline.}
\label{fig:MagicAngle}
\end{figure}

\noindent We receive an areal density of Fe atoms of $8.098~\frac{\text{atoms}}{nm^{2}}$. To calculate the coverage of the sample, we use now the Fe-L$_{3}$ - edge XAS spectrum, which was measured at the VEKMAG beamline for the sub-ML sample at RT with linearly polarized light under the magic angle of 55$^{\circ}$ (see Fig. \ref{fig:MagicAngle}). The measurement was performed with linear polarized light at the magic angle to remove the contribution of different polarizations, since at 55$^{\circ}$ they are equal.
The intensity difference between the L$_{3}$ peak and the pre-edge is around 11\% for the sub-ML sample measured at the VEKMAG beamline and will be defined as the peak height. 
By comparing the areal density which was calculated for Fe(bpz)\textsubscript{2}(phen) by T. G. Gopakumar et al. in Ref. \cite{gopakumar_spin-crossover_2013}, we obtain a conversion factor of 9.87 ($\frac{\text{FeBr\textsubscript{2} $8.10$}}{\text{Fe molecule $0.82$}}$). The Fe L$_{3}$ peak measured for the Fe-based molecule had a peak height of 1.2\% for a 0.8 ML thick sample, which would result in a peak height for 0.8 ML of FeBr\textsubscript{2} of 12\% by using the above conversion factor. Since the measured peak height in our case is lower (11\%), we have an approximated coverage of 0.7 ML.

\noindent This means that the BL sample which was measured at the VEKMAG beamline should be around 2.4 ML thick (evaporation time for the BL was scaled by a factor of 3.4 from the sub-ML data). However, from the thickness calculations we obtain 2.9 ML, which could be related to a different measurement position or additional island growth before finishing the complete layer over the whole sample.

\noindent The calculated thickness is in agreement with the expected thickness by comparing the BOREAS STM (Fig. \ref{fig:BOREAS_STM} (a)) images with the measured LEED image.

\subsection{STM pattern angle calculation}
\begin{figure}[h!]
\centering
\includegraphics[width=\linewidth]{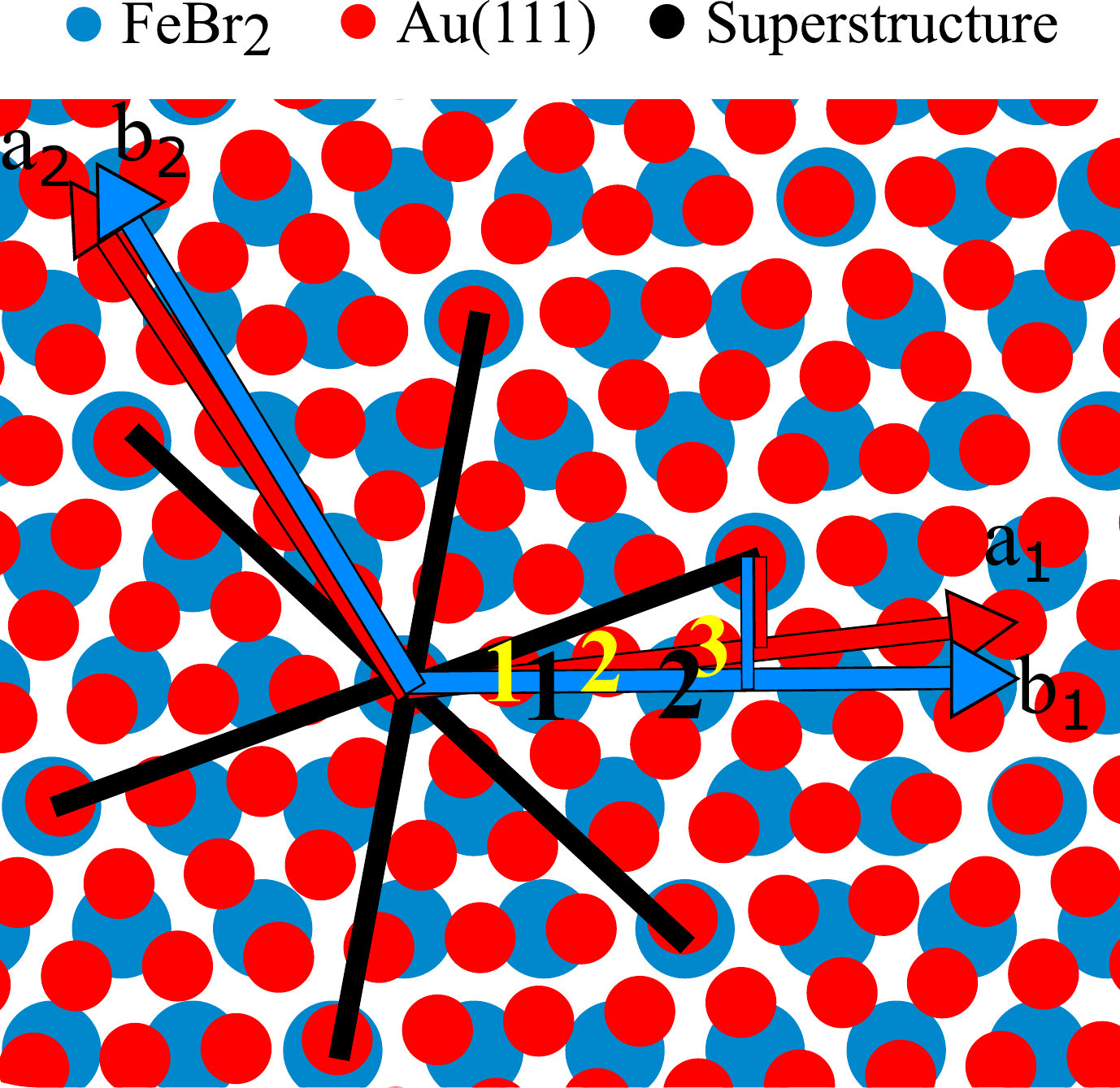}
\caption[Schematic STM]{Overlayed structure of FeBr\textsubscript{2} in blue and Au(111) in red. The black lines are indicating the superstructure hexagon. The red and blue arrows are representing the direction of the substrate and FeBr\textsubscript{2} lattice vectors and are used as a guide to the eye. The yellow and black numbers are referring to the Au atom and Br atom steps to reach a point of the superstructure.}
\label{fig:WKSim}
\end{figure}

\noindent From Fig. \ref{fig:WKSim} we can calculate the vectors, length, and the corresponding angles for the relations between substrate and superstructure as well as between FeBr\textsubscript{2} and superstructure. In Fig. \ref{fig:WKSim} the top-most Br layer is overlayed with the substrate Au(111). In blue the upper Br plane is displayed, which is on top of the Au(111) substrate (red). The black lines are a guide to the eye for the superstructure, which appears at the coincidence points between the Br layer and the Au(111) substrate. 
\noindent The case displayed in Fig. \ref{fig:WKSim} corresponds to the $-5^{\circ}$ case in Fig. \ref{fig:monolayer} (d). In the following the angles and vectors are calculated step by step. The lattice vector of the superstructure in units of the substrate lattice is the following:
\begin{align}
  \vec{a}&=\begin{pmatrix}
    \frac{7}{2} \\
    \frac{\sqrt{3}}{2}
    \end{pmatrix}\\
    \abs{\vec{a}}&=\sqrt{\frac{49}{4}+\frac{3}{4}}=\sqrt{13}\\
    \alpha&=\arctan{\left(\frac{\sqrt{3}}{7}\right)}\approx 13.9^{\circ}
\end{align}
\noindent The superstructure vector is obtained by moving 3.5 times the Au atoms in positive x direction and $\frac{\sqrt{3}}{2}$ in positive y direction (illustrated in Fig. \ref{fig:WKSim} by the red vertical line). 
\noindent For the relation between the FeBr\textsubscript{2} lattice and the superstructure the following relation is obtained.
\begin{align}
  \vec{b}&=\begin{pmatrix}
    \frac{5}{2} \\
    \frac{\sqrt{3}}{2}
    \end{pmatrix}\\
    \abs{\vec{b}}&=\sqrt{\frac{25}{4}+\frac{3}{4}}=\sqrt{7}\\
    \beta&=\arctan{\left(\frac{\sqrt{3}}{5}\right)}\approx 19.1^{\circ} \label{eqref:Calc}
\end{align}
The superstructure vector is obtained by moving 2.5 times the Br-atoms in positive x direction and moving $\frac{\sqrt{3}}{2}$ in positive y direction (illustrated in Fig. \ref{fig:WKSim} blue vertical line).
\noindent To identify now the angle between the substrate and FeBr\textsubscript{2}, the scalar product between the vectors $\vec{a}$ and $\vec{b}$ needs to be calculated.
\begin{align}
  \vec{a}\cdot\vec{b} &=\begin{pmatrix}
    \frac{7}{2} \\
    \frac{\sqrt{3}}{2}
    \end{pmatrix} \cdot \begin{pmatrix}
    \frac{5}{2} \\
    \frac{\sqrt{3}}{2}
    \end{pmatrix}\\
    \abs{\vec{a}}\cdot \abs{\vec{b}} \cdot \cos(\gamma) &=\sqrt{13}\cdot \sqrt{7}\cdot \cos(\gamma)\\
    \gamma&=\arccos{\left(\frac{19}{2\sqrt{91}}\right)}\approx 5.21^{\circ}
\end{align}

\noindent This results in an angle of $5.21^{\circ}$ between the substrate and FeBr\textsubscript{2}. Furthermore a lattice mismatch can be identified, which is causing a strain effect on the FeBr\textsubscript{2}. The expected lattice constant ratio between FeBr\textsubscript{2} and Au(111) (surface lattice constant) would be $\frac{3.776}{2.86}=1.32$. However, the calculations are revealing a ratio of $\sqrt{\frac{13}{7}}=1.36$, which means that the FeBr\textsubscript{2} lattice constant is increased by 3\% with respect to the theoretical lattice value and 6\% to the measured value. This value is still in good agreement with the expected ratio and also within the error range of the measured value.

\subsection{STM - Interatomic distances}
\begin{figure}[h!]
\centering
\includegraphics[width=0.65\linewidth]{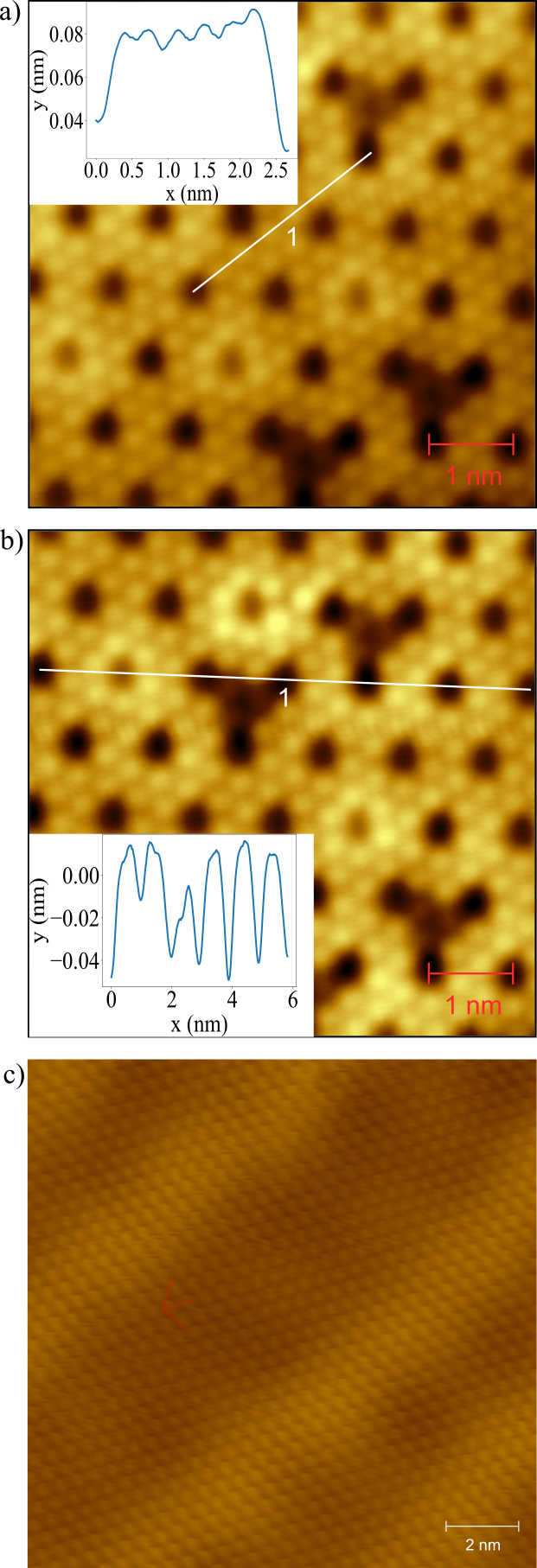}
\caption[Interatomic distance]{Atomic-resolution LT-STM images. In a) the atomic-resolution STM image used to determine the FeBr\textsubscript{2} lattice distance is displayed and the resulting line profile from Line 1 is shown as an insert. In b) the STM image for the superstructure distance is displayed with the inserted line profile. The measurement parameters are U$_{\text{Bias}}=2\cdot 10^{-3}$ V and I$_{\text{TC}}=0.39\cdot 10^{-9}$ A. In c), the atomic-resolution image of Au(111) is displayed, which was measured in the same setup as FeBr\textsubscript{2}. The measured interatomic distance for Au(111) is $2.78 \text{ \AA}$, which is around 5\% smaller than the expected one for Au(111) with $2.86 \text{ \AA}$. The lattice constant was calculated by measuring the interatomic distance of the Au atoms along the marked red line in c) and calculating the average value. The STM measurement was performed at U$_{\text{Bias}}=1\cdot 10^{-3}$ V and I$_{\text{TC}}=1\cdot 10^{-9}$ A.} 
\label{fig:InteratomicLines}
\end{figure}

\noindent In Fig. \ref{fig:InteratomicLines}, examplary images from which we calculated the interatomic distance of FeBr\textsubscript{2} (Fig. \ref{fig:InteratomicLines} (a)) and the superstructure (Fig. \ref{fig:InteratomicLines} (b)) are shown. The used software was Gwyddion. For calculating the interatomic distance, we draw a line between two spots of the superstructure. The drawn line covers 6 atoms. The resulting line profile (\ref{fig:InteratomicLines} a) and b) insert) is used to determine the distances.
\noindent The interatomic distance was obtained as the average over 50 line profiles. As a result we obtained an average value with standard deviation of $3.66\pm 0.04 \text{ \AA}$. As a systematic error of the measurements we assume a relative error of 7\% ($0.26 \text{ \AA}$). This error is based on the atomic-resolution Au(111) STM file (see Fig. \ref{fig:InteratomicLines} (c)). The herringbone reconstruction is also visible in Fig. \ref{fig:InteratomicLines} (c) with the background color.
The superstructure distance was calculated as the average over 35 line profiles. As a result we obtained a value of $9.69\pm 0.05 \text{ \AA}$. As a systematic error of the measurements we assume a relative error of 7\% (resulting in an error of $0.67 \text{ \AA}$).

\newpage
\subsection{STM measurements at the BOREAS beamline}
\begin{figure}[h!]
\centering
\includegraphics[width=\linewidth]{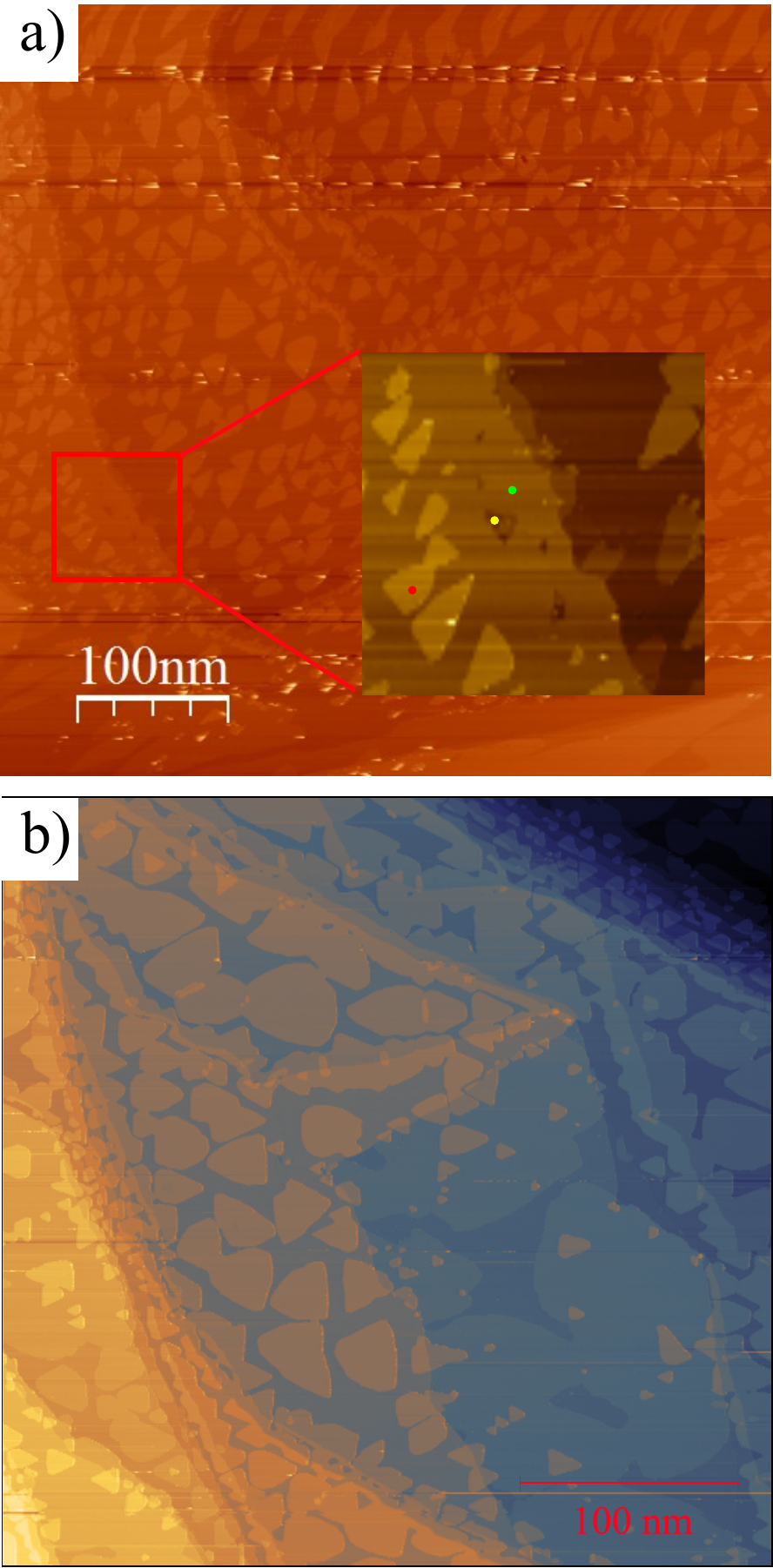}
\caption[LN2-STM BOREAS]{a) STM topography scans of the 1.5 ML sample (U$_{\text{Bias}}=1$ V and I$_{\text{TC}}=0.01\cdot 10^{-9}$ A) measured at 77 K. The red rectangle shows the indication for the ML growth. The insert shows the zoomed-in region with the ML growth indication. The images were evaluated by WSxM \cite{horcas_textlessspan_2007}. b) STM topography scans of the 2.0-ML sample at 77~K (U$_{\text{Bias}}=1$~V and I$_{\text{TC}}=0.02\cdot 10^{-9}$~A). The yellow, green, and red dots are indicating the substrate, the first layer, and the second layer inside the insert.}
\label{fig:BOREAS_STM}
\end{figure}

\noindent The STM measurements were performed directly at the BOREAS beamline at 77~K before measuring XAS and XMCD.
In Fig. \ref{fig:BOREAS_STM}, the evaporated 1.5-ML sample shows the start of the growth of the second and third layer (triangular-shaped islands) on top of the ML. We can also observe carpeting effects for the different heights.
The red rectangle indicates the region of a defect, where we can see the layer underneath of the ML sample. 
\noindent From the STM measurements it is visible that with increasing thickness the top layers grow simultaneously with the completion of the lower layers.

\subsection{STS measurements}
\begin{figure}[h!]
\centering
\includegraphics[width=\linewidth]{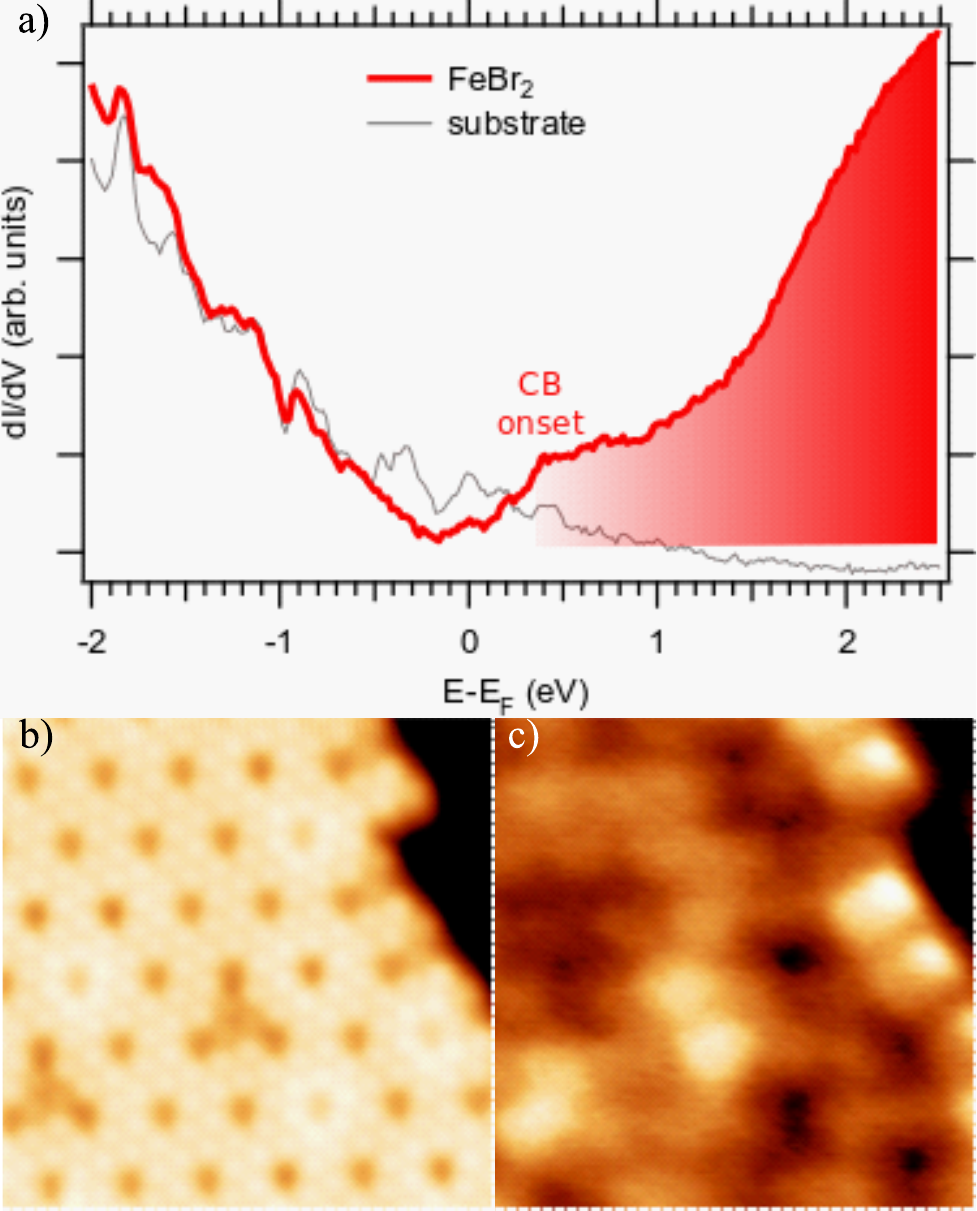}
\caption[STS]{a) STS dI/dV spectra on the FeBr\textsubscript{2} sub - ML (red line) along with the reference spectrum on the substrate (grey line), revealing the conduction band and its associated onset. b) Constant-current “topographic” image (6.1~nm x 6.1~nm) of the FeBr\textsubscript{2} sub - ML and c) a simultaneously acquired conductance map taken at the energy of the CB onset (0.4~eV), proving its distribution over the whole layer while being absent on the bare substrate regions (top-right corner). The tunneling parameters are U$_{\text{Bias}}=0.4$~V and I$_{\text{TC}}=0.05\cdot 10^{-9}$~A.}
\label{fig:STS}
\end{figure}

\noindent The STS measurements in Fig. \ref{fig:STS}  were performed at 4.3~K on the same sub-ML sample as displayed in Fig. \ref{fig:monolayer} (a-b). From the measured dI/dV spectra it was not possible to determine the bandgap. The data reveals that the sub-ML sample is semiconducting with a CB onset at 0.4~eV with respect to the Fermi level. 

\begin{figure*}[ht!]
\centering
\includegraphics[width=\textwidth]{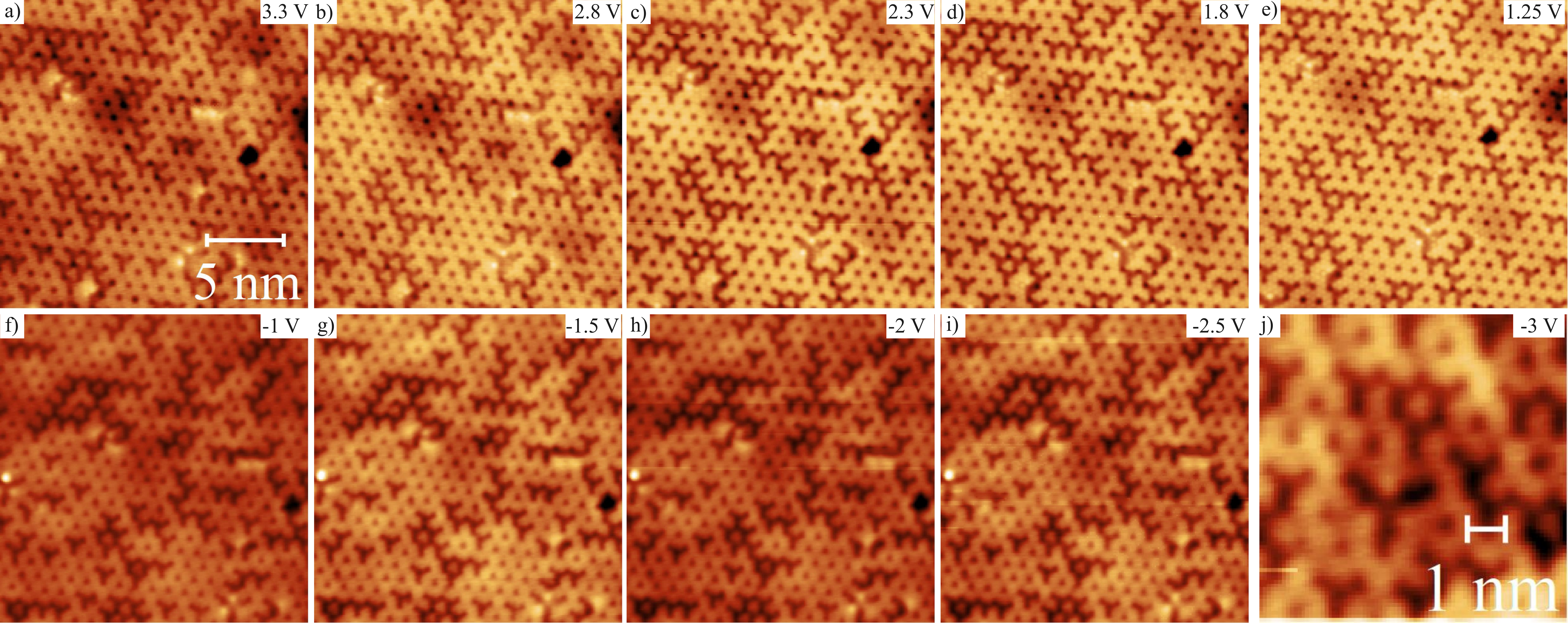}
\caption[Bias dependence]{STM images of a sub-ML sample of FeBr\textsubscript{2} on Au(111) measured at different bias voltages (from 3.3~V to -3~V). The images except j) are showing about the same sample position for all measurements from 3.3~V down to -2.5~V. The tunneling current is I$_{\text{TC}}=0.1\cdot 10^{-9}$~A.}
\label{fig:BiasDep}
\end{figure*}

\noindent To investigate if the superstructure is the result of the electronic properties of the first layer, a bias-dependent measurement series was performed (see Fig. \ref{fig:BiasDep}). The bias voltage was varied from 3.3~V to -3~V. For all different bias voltages only a weak contrast change of the superstructure pattern has been observed. The observed superstructure seems to be bias-independent and therefore can not be of a purely electronic origin. All images besides Fig. \ref{fig:BiasDep} j) are showing the same area of the sample and have the same size.
\newpage
\subsection{Second and third layer of FeBr\textsubscript{2} on Au(111)}
\begin{figure*}[ht!]
\centering
\includegraphics[width=\textwidth]{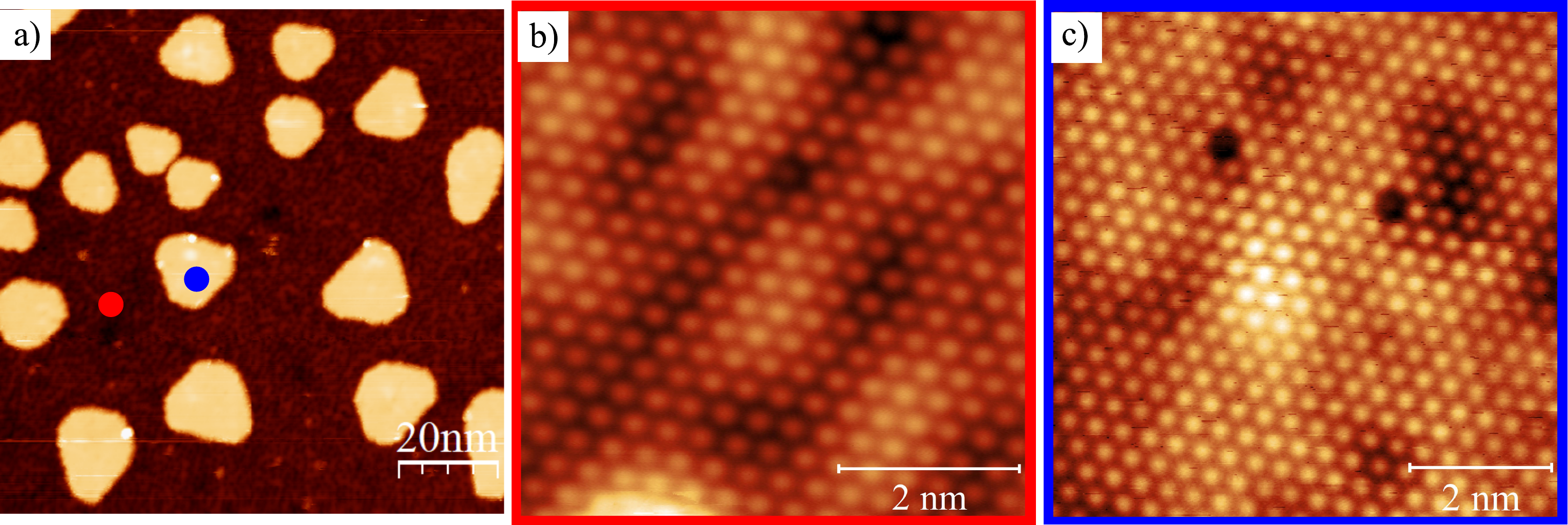}
\caption[Second and Third layer]{STM images of a 2.2-ML sample. a) Overview image. Here the complete second layer and the start of the growth of the third layer (triangular islands) can be observed. b) and c) show atomic-resolution images of the second and third layer, respectively. For the second layer some corrugation is shining through. a) was measured at U$_{\text{Bias}}$=1.7 V and I$_{\text{TC}}=0.04\cdot 10^{-9}$~A, b) at U$_{\text{Bias}}$=-0.2 V and I$_{\text{TC}}=0.4\cdot 10^{-9}$~A, and c) at U$_{\text{Bias}}$=1.6 V and I$_{\text{TC}}=0.1\cdot 10^{-9}$~A. The second and third layer are indicated by red and blue dots in a). The corresponding atomic resolution images have a red and blue frame for the second and third layer.}
\label{fig:BLTL}
\end{figure*}

\noindent Fig. \ref{fig:BLTL} shows STM images of the growth of the second and third layer of FeBr\textsubscript{2} on Au(111). The third layer grows in triangular islands on top of the second layer (Fig. \ref{fig:BLTL} a)). The total coverage of the sample is around 2.3 ML. In b) and c)  atomic-resolution images of the second and third layer are displayed, here no superstructure is visible anymore and just a regular hexagonal pattern is observed. The extracted lattice constants for the second and third layer are in good agreement with the literature values for the bulk sample. For the second layer a lattice constant of $3.78 \pm 0.26\text{ \AA}$ and for the third layer a lattice constant of  $3.75 \pm 0.26\text{ \AA}$ were measured. The second layer shows certain corrugation of the surface, but it has different symmetry and periodicity than the reconstruction in the first layer and probably  could be stress-related. This corrugation is already weaker in the third layer.

\clearpage
\subsection{XPS - Fit parameters for sub - ML and BL samples}
\noindent \noindent In table \ref{ref:XPS} the fit parameters for the sub-ML and BL sample are shown. The samples were grown in the same chamber under the same conditions. After preparation of the sub-ML sample, the substrate was sputtered and annealed. The XPS measurements were performed by using a pass energy of 30~eV with the lens setting of medium area at 1.5~kV and analyzer workfunction of 4.309~eV. The Al-anode with an excitation energy of 1486.61~eV was used and the analyser has an energy resolution of 0.1~eV.

\begin{table}[h!]
\centering
\begin{tabular}{c*{6}{c}c}
Coverage  & Element-Peaks & FWHM (eV)  & Peak position (eV)\\
\hline
\multirow{2}{*}{sub-ML} & Br 3p 3/2 & 3.01$\pm$0.05 & 182.34$\pm$0.03\\
 & 3p 1/2 &   3.025$\pm$0.07& 188.98$\pm$0.04\\
 \hline
\multirow{2}{*}{sub-ML} & Au 4d 5/2 & 5.73$\pm$0.18&353.16$\pm$0.03\\
 & 4d 3/2 &  5.66$\pm$0.03 & 335.25$\pm$0.02\\
 \hline
\multirow{4}{*}{sub-ML} & Fe 2p 3/2 & 3.59$\pm$0.11 & 709.38$\pm$0.06\\
 & Fe 2p 1/2 &  3.78$\pm$ 0.21&722.64$\pm$0.11\\

 & Fe Sat. 2p 3/2  & 7.10$\pm$0.31 & 714.57$\pm$0.16\\

 & Fe Sat. 2p 1/2  & 7.20$\pm$ 0.40&728.87$\pm$0.26\\
\hline
\multirow{2}{*}{BL} & Br 3p 3/2 & 2.78$\pm$0.02 & 182.69$\pm$0.01\\
 & 3p 1/2 &   2.81$\pm$0.04& 189.30$\pm$0.02\\
 \hline
\multirow{2}{*}{BL} & Au 4d 5/2 & 5.510$\pm$0.004&353.18$\pm$0.03\\
 & 4d 3/2 &  5.48$\pm$0.03 & 335.27$\pm$0.02\\
 \hline
\multirow{2}{*}{BL} & Fe 2p 3/2 &3.60$\pm$0.09 & 709.87$\pm$0.03\\

 & Fe 2p 1/2 &  3.82$\pm$0.09 & 723.18$\pm$0.05\\

 & Fe Sat. 2p 3/2  &  6.92$\pm$0.13& 715.13$\pm$0.07\\

 & Fe Sat. 2p 1/2  & 6.61$\pm$0.18& 729.38$\pm$0.08\\
 \hline
\end{tabular}
\caption[XPS-spectra]{Element - specific peak positions and their full width half maximum (FWHM) including the respective error, which were obtained from the fitting routine (lmfit \cite{newville_lmfit_2014}).}
\label{ref:XPS}
\end{table}

\noindent The number of scans was kept constant for each element.
\noindent To check the sample thickness and how it changed the attenuation, we used the element-specific maximum of the Fe 2p3/2 -peak and calculated the height to the post-edge. We calculated the height for the background-subtracted and raw data. For both data sets the same result was obtained. The BL sample shows a stronger Fe and Br signal and a weakened Au signal. In table \ref{ref:XPSratio} we see that for the thicker sample the Au peak is more attenuated. The BL sample needed to be shifted by 0.2~eV to negative binding energies to match the peak position of the sub-ML sample.

\begin{table}[h!]
\centering
\begin{tabular}{c*{6}{c}c}
Coverage  & Element-Peaks & Height\\
\hline
\multirow{2}{*}{sub-ML} & Br & 440,73\\
 & Fe &   612.68\\
 & Au &   11760.62\\
 \hline
\multirow{2}{*}{BL} & Br & 1004.30\\
 & Fe &   1152.93\\
 & Au &   9227.65\\
 \hline
\end{tabular}
\caption[XPS-thickness dependent ratio]{Calculated height of the main peak to the pre-edge for the sub-ML and BL samples of the background subtracted data.}
\label{ref:XPSratio}
\end{table}
\noindent The corresponding LEED images for the sub-ML and BL sample are shown in Fig. \ref{fig:multilayer} (b-c). The peak height is not related to the atomic ratio of Fe and Br of FeBr\textsubscript{2}. The chemical stoichiometry stays for both samples the same. The theoretical area ratio for a 2p core level peak is 2 and we receive from the fit for the sub-ML 1.9 and for the BL 1.7. The difference is caused by the not fitted multiplet splitting of the Fe$^{2+}$ HS state, because we do not have the energy resolution to fit them. The satellite FWHM is around 1.8 times bigger than the core-level FWHM and is in good correspondence to \cite{martin-garcia_unconventional_2016}. The Fe to Br ratio was checked by using the area of the Fe main peaks (excluding the satellite peaks) and the Br main peaks. 
\begin{align}
R=\frac{A_{Fe2p}/S_{Fe2p}}{A_{Br3p}/S_{Br3p}}
\end{align}
\noindent R is the ratio between Fe and Br and S is the element specific sensitivity factor (atomic sensitivity factor) for an angle of $55^{\circ}$ ($S_{Fe2p}=2.957$ and $S_{Br3p}=1.279$) \cite{biesinger_relative_2009,moulder_handbook_1992}. As a result we obtained for the sub-ML sample a ratio of 0.56 and for the BL sample a ratio of 0.58. This is in good agreement with the expected ratio of 0.5 for FeBr\textsubscript{2}. By using a different sensitivity factor from the Wagner paper, the values are slightly different for the bulk approximation (R=0.49) and strongly different for the surface approximation (R=0.34) \cite{wagner_sensitivity_1983}. The surface approximation is not valid for the measured XPS because the resolution is not good enough to distinguish between surface and bulk peaks and the bulk phase is the more dominant one. For this sample, a ML consists of two Br planes sandwiching the transition-metal plane (Fe). In the case of using the sensitivity factors from Ref. \cite{shard_intensity_2019}, we obtain a ratio of 0.61. The calculated sensitivity factors from that paper have all a 10\% error. Also the value of 0.61 as a ratio is in good agreement with the expected value of 0.5 for FeBr\textsubscript{2}. Another contribution to the not perfect ratio of 0.61 could be the use of a non-monochromatic x-ray gun, which results in doublet peaks where the extra peaks are displaced towards lower binding energies by 9.8~eV with an intensity percentage of 6.4\% of the real peaks.

\subsection{Sample thickness approximated by the integrated averaged XAS spectra}
\noindent To calibrate the sample thicknesses, we are using the value of the integrated isotropic XAS spectra normalized to the pre-edge (r value in the Table\ref{tab:sum_rules_full}), as well as the average XAS peak height (PH$=max(\frac{\sigma^{+}+\sigma^{-}}{2})$). The reference sample for the thickness calibration is a 0.7 ML sample (Fig. \ref{fig:MagicAngle}). The thickness is calculated as the average of both geometries (NI and GI). A ML sample has an r-factor of 0.5 and an PH value of 0.2. The only sample for which we do not have a nearly equal signal for the r-factor and the XAS height is the 2.9 ML one. The 2.9 ML was calculated from the NI signal, for GI we obtain a sample thickness of 5.0 ML. The reason could be that the sample intensity at GI was measuring a thicker amount of the material. Due to the island growth it could have created thicker regions which were measured at GI. 
For the VEKMAG measurements we used a focus/beam spot size of 0.8 mm $\times$ 0.8~mm. The error for the calculated coverages is approximated by comparing the NI and GI values. 
Besides the BL, which was measured at BESSY, the systematic error is around $\pm 0.4$ ML, which is due to uncertainties and observations from STM.

\begin{table}[h!]
\begin{tabular}{cccccc}
C (ML) & Geometry & Beamline &T (K) &  PH (a.u.) &r (a.u.)\\
\hline
\multirow{2}{*}{0.7}  & NI & \multirow{4}{*}{VEKMAG} & \multirow{4}{*}{10} & 0.14& 0.38  \\ 

 &GI &  & & 0.15& 0.41  \\ \cline{1-1} \cline{2-2} \cline{5-5} \cline{6-6}  

\multirow{2}{*}{2.9}  &NI &  & & 0.58& 1.45 \\

  &GI &  &  & 0.94& 2.58  \\
\hline
\multirow{2}{*}{0.6}  &NI & \multirow{6}{*}{BOREAS} & \multirow{6}{*}{2} & 0.12& 0.28\\ 

 & GI &  & & 0.12& 0.28  \\ \cline{1-1} \cline{2-2} \cline{5-5} \cline{6-6} 

\multirow{2}{*}{1.5} & NI &  & & 0.30& 0.74  \\

  &GI & &  & 0.31& 0.81 \\ \cline{1-1} \cline{2-2} \cline{5-5} \cline{6-6} 

\multirow{2}{*}{2.0}  &NI &  &  & 0.40& 0.98 \\

  &GI & & & 0.44& 1.03\\
\hline
\end{tabular}
\caption[Whiteline and XAS height]{Calculated whiteline height values (r) and the average XAS peak height (PH) for the XAS spectra after the average, normalization and background correction for different FeBr\textsubscript{2} thicknesses measured at different geometries. The conversion factor is 0.02 PH=0.1 ML (0.7 ML is equal to 0.14 PH).}
\label{tab:sum_rules_full}
\end{table}

\subsection{XAS and XMCD shift corrections}
\begin{figure}[h!]
\centering
\includegraphics[width=\linewidth]{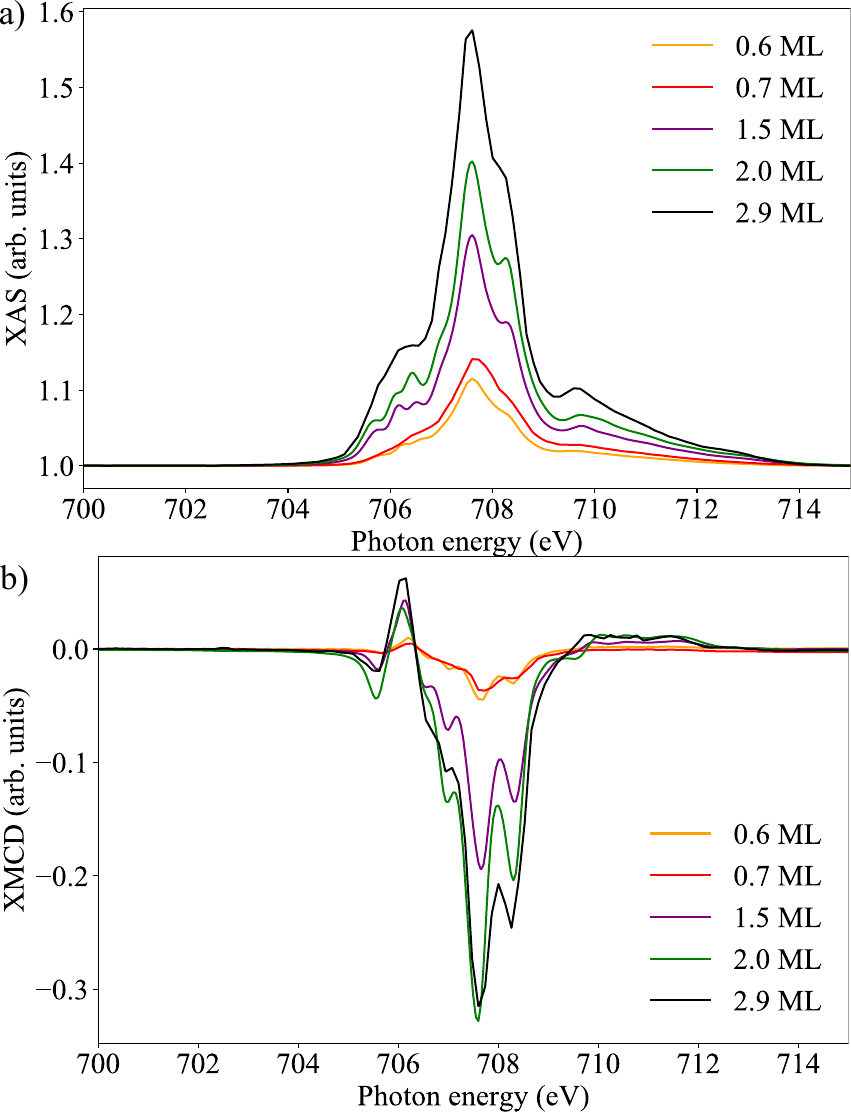}
\caption[Different thickness XAS and XMCD spectra]{In panel a) the XAS spectra for the different thicknesses measured at 6~T with circular polarization are displayed. The 0.7-ML and 2.9-ML samples were measured at the VEKMAG beamline at BESSY II at 10~K and the other thicknesses (0.6, 1.5 and 2.0 ML) were measured at the BOREAS beamline at ALBA at 2~K. Panel b) shows the corresponding XMCD spectra.}
\label{fig:ALBABESSYXMCDXAS}
\end{figure}

\noindent In Fig. \ref{fig:ALBABESSYXMCDXAS} the beamline-dependent energy-corrected XAS and XMCD spectra are displayed. To overlay the BESSY spectra with our ALBA data, we shifted the spectra of the 0.7-ML and 2.9-ML samples by around $1.6$~eV. The comparison signal to align the data is the ALBA sub-ML signal. The BESSY data were acquired at 10~K and the ALBA data at 2~K. The used measurement point density at ALBA is around 3 times bigger than the one at BESSY. All the measurements took place at NI. The fact that the XMCD signals for the 2.0- and 2.9-ML samples are equal could be caused by the temperature difference. 
\noindent The BOREAS data were also energy-corrected with the sub-ML spectra as a reference. The 2.0-ML sample was corrected by 0.42~eV and the 1.5-ML one by 0.61~eV to higher energies. This shifting could be caused by two different effects. On the one hand, since the thinnest sample (sub-ML) was measured in a different experiment, one year before measuring the rest, the beamline energy calibration may have changed during this time. On the other hand, while increasing the thickness, the material becomes more insulating, changing the gap and therefore the final states. The reason for the shifts is the monochromator movement and also the workfunction change in TEY.
\begin{figure}[h!]
\centering
\includegraphics[width=\linewidth]{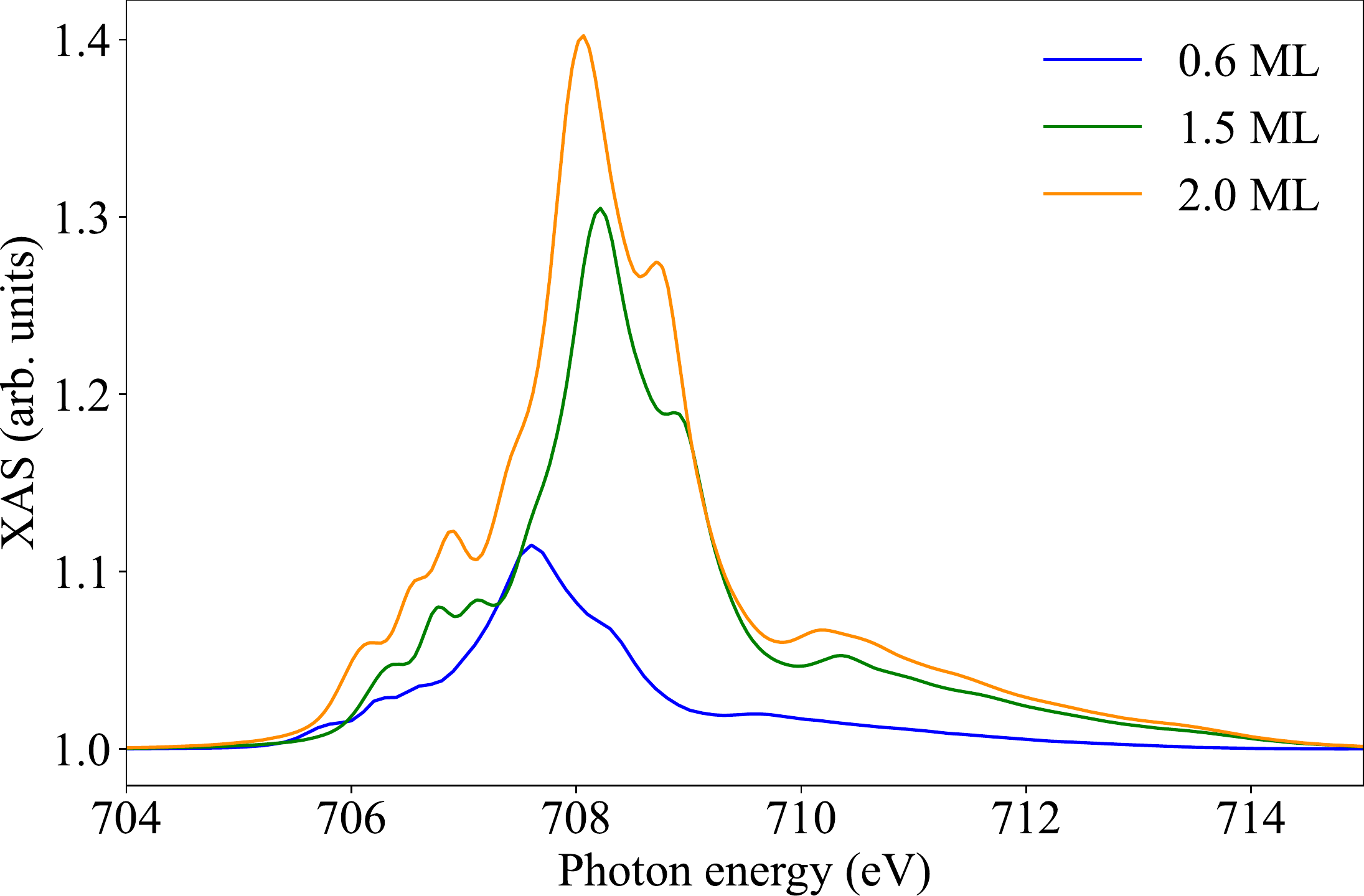}
\caption[Shift corrected spectra]{Shown are the XAS spectra before shift correction taken at the BOREAS beamline at ALBA at 2~K and NI.}
\label{fig:ShiftCorrect}
\end{figure}

\subsection{XAS - Fe$^{2+}$}
\begin{figure}[h!]
\centering
\includegraphics[width=\linewidth]{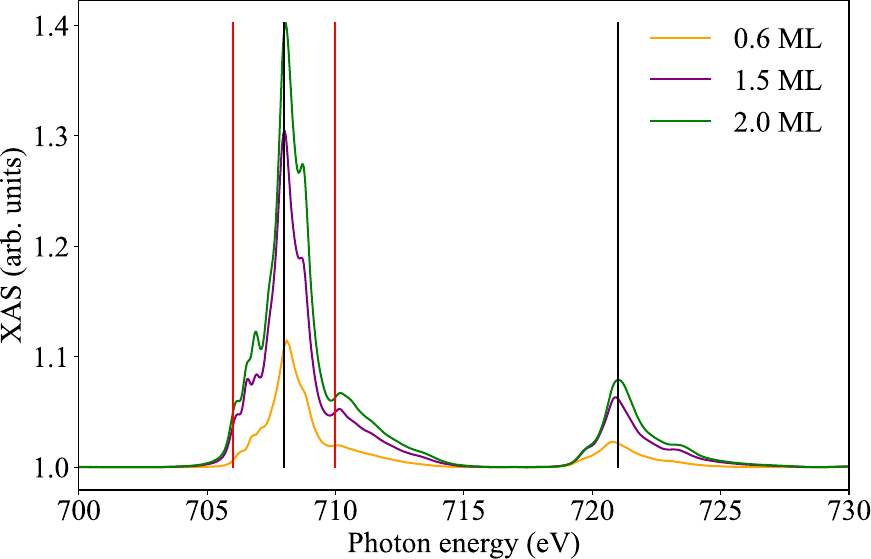}
\caption[Fe2+ state]{Displayed are the averaged XAS spectra for the different thicknesses measured at 2~K and 6~T. The vertical lines are used as a guide to the eye to compare L$_{3}$ and L$_{2}$ edge position (black lines) with the multiplet peak positions (red lines). The displayed spectra for the different coverages were all measured at the BOREAS beamline at ALBA.}
\label{fig:Fe2+}
\end{figure}

\noindent The sample only shows a single stoichiometry (Fe$^{2+}$). In Fig. \ref{fig:Fe2+} the averaged signals of the 0.6-ML, 1.5-ML and 2.0-ML samples are shown (full range L$_{3}$ and L$_{2}$-edge). 

\noindent By averaging both polarizations $\sigma^{+}$ and $\sigma^{-}$, the XAS spectra without magnetic contributions can be calculated. Fig. \ref{fig:Fe2+} shows the shift-corrected data for the different thicknesses and the black lines indicate the position of the main peaks (L$_{3}$ and L$_{2}$ edge). The red lines represent the side peaks which are known for an Fe$^{2+}$ state with octahedral symmetry \cite{koehl_detection_2013,gopakumar_spin-crossover_2013,gimenez-marques_graftfast_2018,gota_characterization_1997,sassi_first-principles_2017}. In Fig. \ref{fig:Fe2+} the shift-corrected BOREAS measurements (0.6 to 2.0 ML) are displayed.

\subsection{Sum-rule analysis of the XMCD data} \label{subsec:SI_XMCD}

\noindent The values for the effective spin and orbital magnetic moments were obtained from the areas of the $L_{3}$ and $L_{2}$ peaks of the XMCD spectra from the sum rules \cite{kuch_x-ray_2004, chen_experimental_1995, carra_x-ray_1993}:
\begin{align}
m_{s, eff}&=-\frac{A_{L_3}-2\cdot A_{L_2}}{A_\text{Average}}\cdot N_{h}\cdot \frac{1}{\sigma}\label{ref:mseff}\\
m_{l}&=-\frac{2}{3}\cdot\frac{A_{L_3}+\cdot A_{L_2}}{A_\text{Average}}\cdot N_{h}\cdot\frac{1}{\sigma}
\end{align}
where $N_{h}$ is the number of holes (= 4 for Fe$^{2+}$ in FeBr\textsubscript{2}), $A_{L_3}$, $A_{L_2}$ and $A_\text{Average}$ are the areas of the XMCD $L_{3}$ and $L_{2}$ region and the total area of the isotropic XAS. $\sigma$ represents the degree of circular polarization, which is beamline-dependent. 
The $m_{s,eff}$ value calculated via equation \eqref{ref:mseff} is the effective spin magnetic moment, result of the sum of the actual spin moment plus the magnetic dipole term, $\frac{7}{2}T_{Z}$. The background correction was performed using an asymmetrically reweighted penalized least- squares smoothing \cite{baek_baseline_2015}.

\noindent In table \ref{Tab:MagMom} the magnetic moment values obtained from the sum-rule analysis for the different samples are displayed. The BOREAS data were measured by using a 6~T field and switching the polarization and the VEKMAG data by keeping the polarization fixed and ramping the field ($\pm$ 6~T).

\begin{table}[h!] \label{tab:sum_rules_full}
\centering
\begin{tabular}{cccccccc}
 \hline
\multicolumn{1}{c|}{\multirow{3}{*}{C (ML)}} & \multicolumn{1}{c|}{\multirow{3}{*}{T (K)}} & \multicolumn{6}{c}{$\mu$ ($\mu_B$/Fe at)}                            \\ \cline{3-8} 
\multicolumn{1}{c|}{}                   & \multicolumn{1}{c|}{}                   & \multicolumn{3}{c|}{NI}        & \multicolumn{3}{c}{GI} \\ \cline{3-8} 
\multicolumn{1}{c|}{}                   & \multicolumn{1}{c|}{}                   & \multicolumn{1}{c|}{$m_{\text{s eff}}$} & \multicolumn{1}{c|}{$m_{l}$} &\multicolumn{1}{c|}{R ($\frac{m_s}{m_l}$)}& \multicolumn{1}{c|}{$m_{\text{s eff}}$}        & \multicolumn{1}{c|}{$m_{l}$}&\multicolumn{1}{c|}{R ($\frac{m_s}{m_l}$)}\\ \hline
0.6& 2 &1.13 &0.30 &3.8 &1.05 &0.36& 2.9 \\
 \hline
\multirow{6}{*}{0.7}&10 &0.93 &0.25&3.7 &1.00 &0.16& 6.25\\

&12 &-  &-&- &1.25 &0.23 & 5.4\\

&18  &-  &- &- &0.81 &0.33 & 2.5\\

&25 &-  &-&- &0.54 &0.18& 3.0\\

&30 &-  &- &- &0.33 &0.04& 8.3 \\

&35 &-  &- &- & 0.52 &0.11 & 4.7\\
 \hline
1.5&2 &1.814 &0.60&3.0 &2.03 &0.45&4.5\\
 \hline
 \multirow{9}{*}{2.0}&2 &2.15 &0.70 & 3.1 &1.91 &0.47& 4.1\\

&4 &2.05  &0.72 &2.8 &- &-& -\\

&7  &1.97  &0.66& 2.98 &- &-&-\\

&12 &1.87  &0.65& 2.9  &- &-&-\\

&17  &1.69 &0.47& 3.6 &- &-&-\\

&25 &1.23 &0.38 &3.2  &- &-&-\\

&40  &0.88 &0.23 &3.8 &- &-&-\\

&65 &0.54  &0.17& 3.2  &- &-&-\\

&100  &0.39 &0.14& 2.8 &- &-&-\\
 \hline
\multirow{4}{*}{2.9}&10 &2.12 &0.79 & 2.7  &1.21 &0.40& 3.0\\

&12 &1.93  &0.58 &3.3 &1.51 &0.42&3.6\\

&16  &-  &- &-  &1.05 &0.33& 3.2\\

&20 &-  &- &- &1.00 &0.31& 3.2\\
\hline
\end{tabular}
\caption[Magnetic moment for different coverages]{Evaluated magnetic moment values for all the measured coverages at NI and GI for different temperatures.
The used degree of polarization is 77\% for the VEKMAG data (0.7- and 2.9-ML) \cite{prokes_search_2021} and 100\% for the BOREAS data (0.6-, 1.5-, and 2.0-ML). C stands for the coverage in ML, T is the temperature and R is the ratio between the orbital magnetic moment and the effective spin magnetic moment. The error of $m_{\text{s eff}}$ and $m_{l}$ is $\pm$10\%.}
\label{Tab:MagMom}
\end{table}
\newpage

\subsection{Brillouin function for the combined magnetic moment}

\begin{figure}[h!]
\centering
\includegraphics[width=\linewidth]{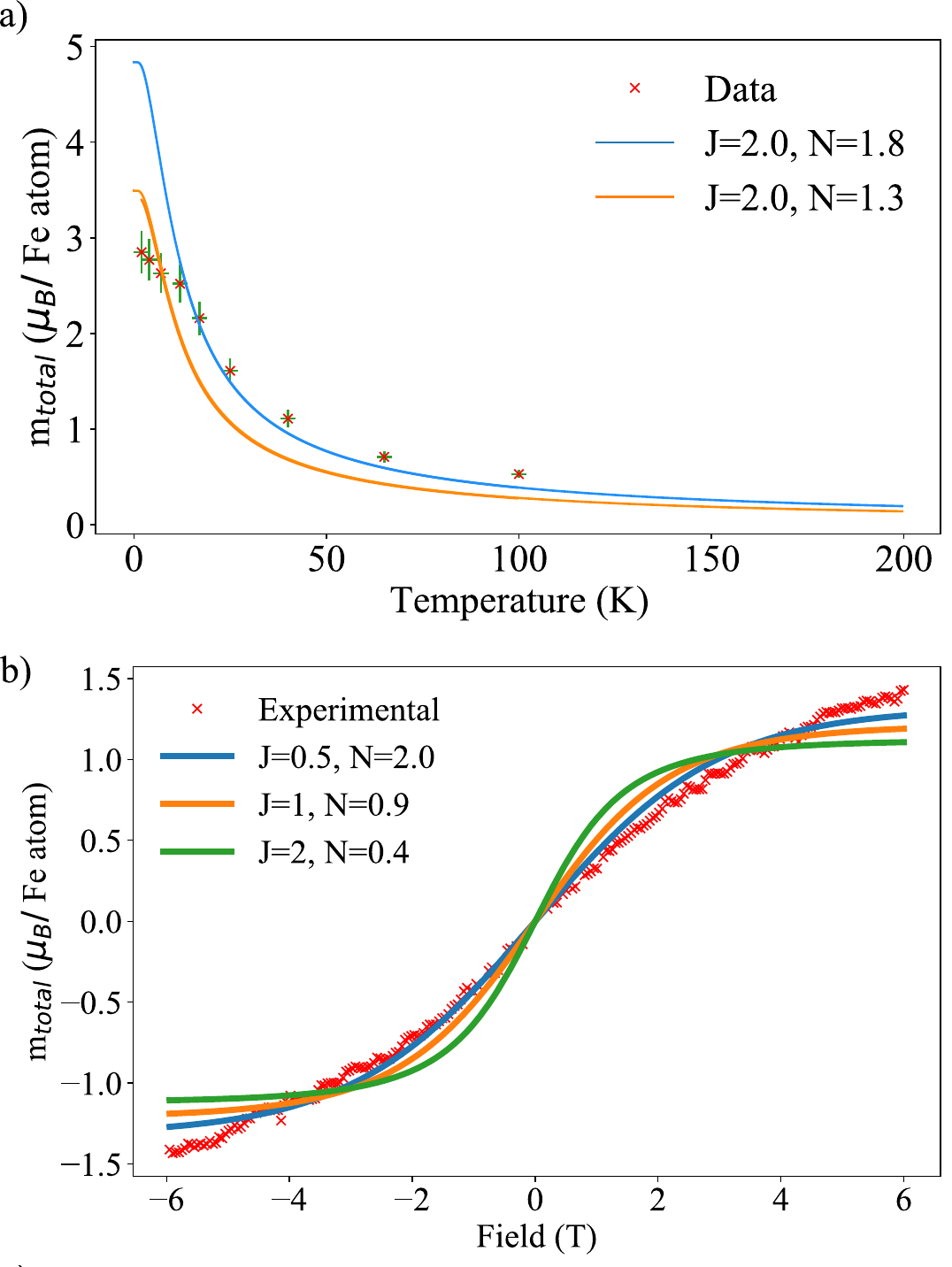}
\caption[Brillouin Fit]{In a) the total magnetic moment for the 2.0-ML sample is compared to the Brillouin function. Two different Brillouin function fits are included. The blue fit is using the data starting at 12~K and orange includes the measurements at lower temperatures (2~K). The magnetic moments were obtained from the measurements performed at $B=6$~T. In b) the 0.6-ML magnetization curve at NI and 2~K is fitted by the Brillouin function. Three different functions were used by ranging J from 0.5 to 2 (fixed temperature of 2~K). The best fitting case is the one of $J=0.5$. However, still the match is not perfect and $J=0.5$ is not reasonable for this material. Therefore the material is not paramagnetic. The magnetic moment values were calculated from the BL data measured at the BOREAS beamline and the sub-ML magnetization curve was measured also at the BOREAS beamline.}
\label{fig:BrillouinFit}
\end{figure}

\noindent In Fig.~\ref{fig:BrillouinFit} (a) the total magnetic moment data is fitted with two Brillouin functions for different temperature ranges. The blue fit was performed by excluding the experimental data at lower temperatures (2-10~K). The orange fit is based on the full temperature range starting from 2~K. We see that the decay of magnetization with temperature is always steeper for the Brillouin function, which points towards ferromagnetic interactions in the experimental data. In b) the measured magnetization curve is plotted together with the simulated Brillouin functions as a function of field for different $J$-values and different scaling factors. Paramagnetic systems reach saturation faster than the experimental curve. The XMCD loops and the Brillouin functions are normalized to the corresponding magnetic moment of the different samples. For the fit we assumed g = 2 and T = 2~K. As a fitting parameter we used J and N. For the temperature-dependent fit we used a constant field of $B=6~T$. The fitting parameter N had the boundaries $1-10$ and J from $0$ to $2$.
The  fit function is:

\begin{equation} \label{eq1} 
\begin{split} 
M&=N\cdot g\cdot J\cdot\biggl(\frac{2\cdot J+1}{2\cdot J}\cdot \coth({\frac{2\cdot J+1}{2\cdot J}\cdot x})\\
&-\frac{1}{2\cdot J}\cdot \coth({\frac{1}{2\cdot J}\cdot x})\biggr),\\
x&=\frac{g\cdot J\cdot \mu_{B}\cdot B}{k_{B}\cdot T}
\end{split}
\end{equation}

\begin{figure}[h!]
\centering
\includegraphics[width=\linewidth]{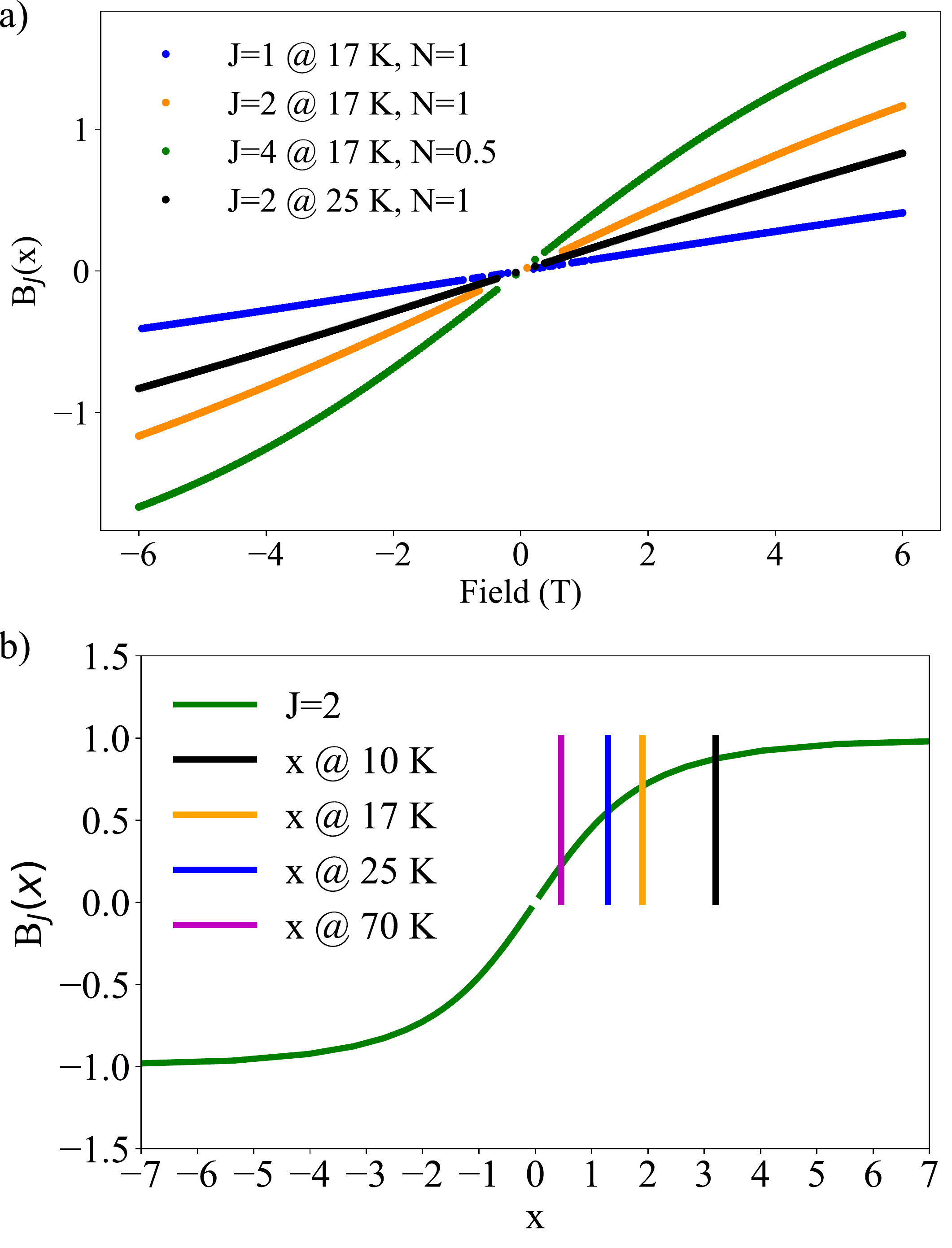}
\caption[Temperature determination for the fit]{a) Brillouin functions for J = 1, J = 2, and J = 4 for a temperature of 17~K as well as for J = 2 and T = 25~K as a function of magnetic field.  b) Brillouin function for J = 2 as a function of x. The vertical lines represent the values of x for a field of 6 T and temperatures of 10, 17, 25, and 70~K.}
\label{fig:BrillX}
\end{figure}

\noindent Fig.~\ref{fig:BrillX} a) shows four exemplary Brillouin functions with J ranging from 1 to 4 at temperatures of 17 and 25 K as a function of field. Except for the curve for J = 4, all curves show approximately linear behavior in the range up to 6 T.  Fig. S13 b) presents the Brillouin function for J = 2 as a function of x.  Vertical bars marked in different colors indicate the x values for a field of 6 T and the indicated temperatures.  Also here it can be seen that at 17 K, the curve is reasonably far from saturation.  Furthermore, as discussed in the main text and shown in Fig. S12, the experimental data are flatter than Brillouin functions, such that the interpretation of the magnetic moment obtained from the XMCD spectra at 6 T in Fig. 5 c) as magnetic susceptibility is justified for T $\geq$ 17~K.

\newpage
\subsection{Br XMCD}
\begin{figure}[h!]
\centering
\includegraphics[width=\linewidth]{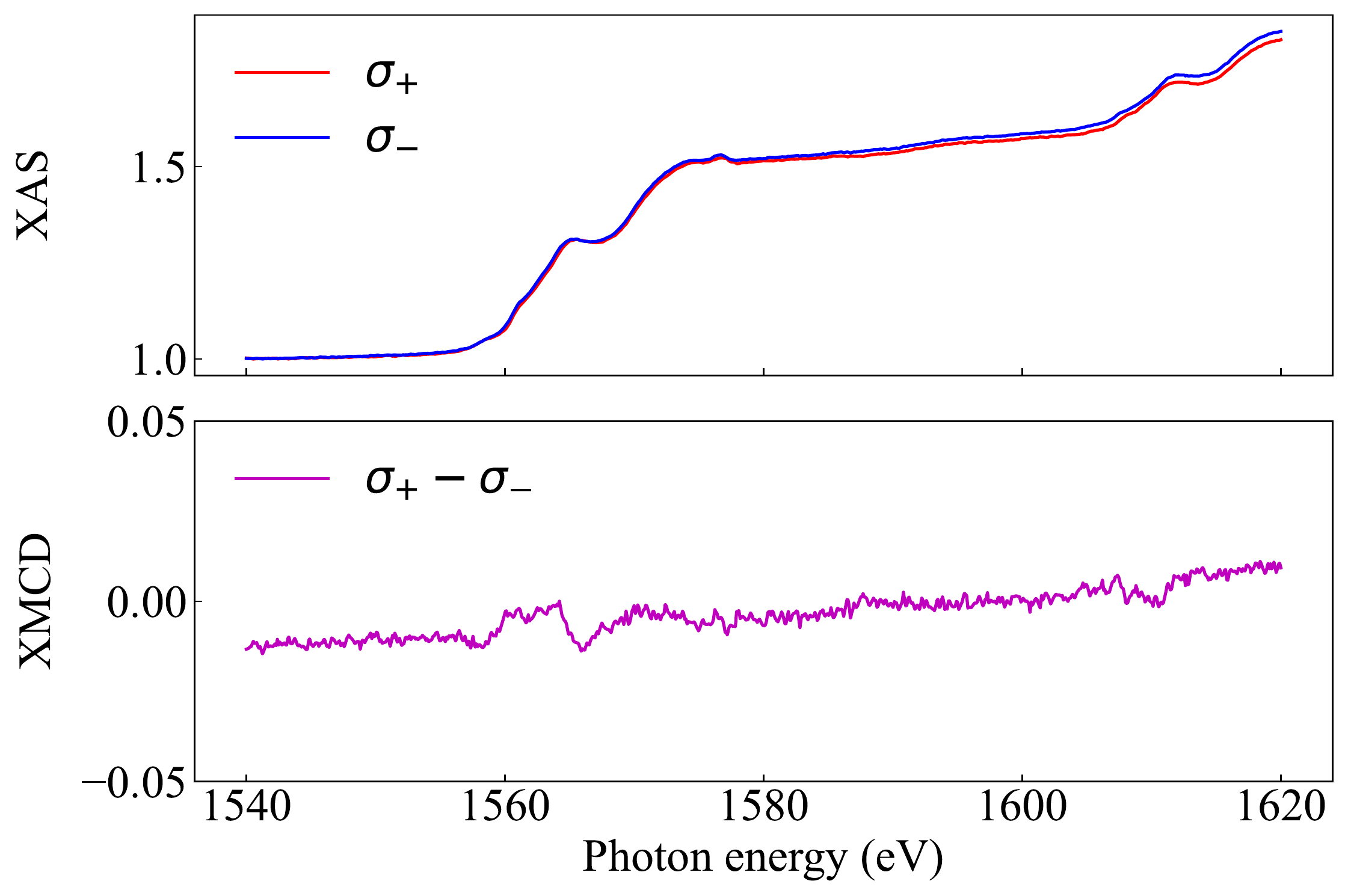}
\caption[XMCD Br L$_{3}$-edge]{XAS and XMCD spectra of the Br L$_{3}$ edge. The measurements were performed on the 1.5-ML sample at 6~T, 2~K and GI (measured at the BOREAS beamline).}
\label{fig:XMCDBr}
\end{figure}

\noindent In Fig. \ref{fig:XMCDBr} the XAS and XMCD measurements at the Br L$_{2,3}$ edge are displayed. The step-like behaviour is a direct consequence of the filled d orbitals. From the XMCD analysis it can be obtained that the magnetic behaviour is not caused by Br.

\clearpage
\section{References}
\bibliographystyle{ieeetr} 
\bibliography{Main}

\end{document}